\documentclass{emulateapj}
\usepackage{apjfonts}
\usepackage{natbib}
\usepackage{graphics}
\usepackage{multirow}
\usepackage{amsbsy}
\usepackage{amsmath}
\usepackage{mathrsfs}
\usepackage[implicit=true, debug=false, 
plainpages=false, colorlinks=true, 
linkcolor=black, urlcolor=black, citecolor=black, filecolor=black, 
hyperindex=true]{hyperref}

\def\bicep{{\sc Bicep}}

\def\dasi{{\sc Dasi}}

\def\spider{{\sc Spider}}

\def\keck{{Keck Array}}

\def\deg{^\circ}

\usepackage{xspace} 
\def\bicepp{{\sc Bicep2}\xspace}
\def\bicep{{\sc Bicep}\xspace}
\def\bicepo{{\sc Bicep1}\xspace}
\def\keck{{\it Keck Array}\xspace}
\def\dasi{{\sc Dasi}\xspace}
\def\quada{{\sc Quad}\xspace}
\def\spider{{\sc Spider}\xspace}
\def\bicepq{{\sc Bicep3}\xspace}

\shorttitle{\bicepp and \keck BEAM CHARACTERIZATION}
\shortauthors{\bicepp and \keck COLLABORATIONS}

\begin{document}

\title{\bicepp / \keck IV: Optical Characterization and Performance of the \bicepp and \keck Experiments}
\author{The \bicepp and \keck Collaborations - P.~A.~R.~Ade\altaffilmark{1}}
\author{R.~W.~Aikin\altaffilmark{2}}
\author{D.~Barkats\altaffilmark{3}}
\author{S.~J.~Benton\altaffilmark{4}}
\author{C.~A.~Bischoff\altaffilmark{5}}
\author{J.~J.~Bock\altaffilmark{2,6}}
\author{K.~J.~Bradford\altaffilmark{5}}
\author{J.~A.~Brevik\altaffilmark{2}}
\author{I.~Buder\altaffilmark{5}}
\author{E.~Bullock\altaffilmark{7}}
\author{C.~D.~Dowell\altaffilmark{6}}
\author{L.~Duband\altaffilmark{8}}
\author{J.~P.~Filippini\altaffilmark{2,9}}
\author{S.~Fliescher\altaffilmark{10}}
\author{S.~R.~Golwala\altaffilmark{2}}
\author{M.~Halpern\altaffilmark{11}}
\author{M.~Hasselfield\altaffilmark{11}}
\author{S.~R.~Hildebrandt\altaffilmark{2,6}}
\author{G.~C.~Hilton\altaffilmark{12}}
\author{H.~Hui\altaffilmark{2}}
\author{K.~D.~Irwin\altaffilmark{13,14,12}}
\author{J.~H.~Kang\altaffilmark{13}}
\author{K.~S.~Karkare\altaffilmark{5}}
\author{J.~P.~Kaufman\altaffilmark{15}}
\author{B.~G.~Keating\altaffilmark{15}}
\author{S.~Kefeli\altaffilmark{2}}
\author{S.~A.~Kernasovskiy\altaffilmark{13}}
\author{J.~M.~Kovac\altaffilmark{5}}
\author{C.~L.~Kuo\altaffilmark{13,14}}
\author{E.~M.~Leitch\altaffilmark{16}}
\author{M.~Lueker\altaffilmark{2}}
\author{K.~G.~Megerian\altaffilmark{6}}
\author{C.~B.~Netterfield\altaffilmark{4}}
\author{H.~T.~Nguyen\altaffilmark{6}}
\author{R.~O'Brient\altaffilmark{6}}
\author{R.~W.~Ogburn~IV\altaffilmark{13,14}}
\author{A.~Orlando\altaffilmark{15}}
\author{C.~Pryke\altaffilmark{10}}
\author{S.~Richter\altaffilmark{5}}
\author{R.~Schwarz\altaffilmark{10}}
\author{C.~D.~Sheehy\altaffilmark{10,16}}
\author{Z.~K.~Staniszewski\altaffilmark{2}}
\author{R.~V.~Sudiwala\altaffilmark{1}}
\author{G.~P.~Teply\altaffilmark{2}}
\author{K.~Thompson\altaffilmark{13}}
\author{J.~E.~Tolan\altaffilmark{13}}
\author{A.~D.~Turner\altaffilmark{6}}
\author{A.~G.~Vieregg\altaffilmark{17,16,*}}
\author{A.~C.~Weber\altaffilmark{6}}
\author{C.~L.~Wong\altaffilmark{5}}
\author{K.~W.~Yoon\altaffilmark{13,14}}

\altaffiltext{1}{School of Physics and Astronomy, Cardiff University, Cardiff, CF24 3AA, UK}
\altaffiltext{2}{Department of Physics, California Institute of Technology, Pasadena, CA 91125, USA}
\altaffiltext{3}{Joint ALMA Observatory, ESO, Santiago, Chile}
\altaffiltext{4}{Department of Physics, University of Toronto, Toronto, ON, Canada}
\altaffiltext{5}{Harvard-Smithsonian Center for Astrophysics, 60 Garden Street MS 42, Cambridge, MA 02138, USA}
\altaffiltext{6}{Jet Propulsion Laboratory, Pasadena, CA 91109, USA}
\altaffiltext{7}{Minnesota Institute for Astrophysics, University of Minnesota, Minneapolis, MN 55455, USA}
\altaffiltext{8}{Universit\'{e} Grenoble Alpes, CEA INAC-SBT, F-38000 Grenoble, France}
\altaffiltext{9}{Department of Physics, University of Illinois at Urbana-Champaign, Urbana, IL 61820, USA}
\altaffiltext{10}{Department of Physics, University of Minnesota, Minneapolis, MN 55455, USA}
\altaffiltext{11}{Department of Physics and Astronomy, University of British Columbia, Vancouver, BC, Canada}
\altaffiltext{12}{National Institute of Standards and Technology, Boulder, CO 80305, USA}
\altaffiltext{13}{Department of Physics, Stanford University, Stanford, CA 94305, USA}
\altaffiltext{14}{Kavli Institute for Particle Astrophysics and Cosmology, SLAC National Accelerator Laboratory, 2575 Sand Hill Rd, Menlo Park, CA 94025, USA}
\altaffiltext{15}{Department of Physics, University of California at San Diego, La Jolla, CA 92093, USA}
\altaffiltext{16}{Kavli Institute for Cosmological Physics, University of Chicago, Chicago, IL 60637, USA}
\altaffiltext{17}{Department of Physics, Enrico Fermi Institute, University of Chicago, Chicago, IL 60637, USA}
\altaffiltext{*}{Corresponding author: avieregg@kicp.uchicago.edu}

\begin{abstract}
\bicepp and the \keck are polarization-sensitive microwave telescopes that observe the 
cosmic microwave background (CMB) from the South Pole at degree 
angular scales in search of a signature of inflation imprinted as B-mode polarization in the CMB. 
\bicepp was deployed in late 2009, observed for three years until the end 
of 2012 at 150~GHz with 512 antenna-coupled transition edge sensor bolometers, 
and has reported a detection of B-mode polarization on degree angular scales. The \keck
was first deployed in late 2010 and will observe through 2016 with five receivers at several frequencies 
(95, 150, and 220~GHz).
\bicepp and the \keck share a common optical design and
employ the field-proven \bicepo strategy of using small-aperture, cold, on-axis refractive optics, 
providing excellent control of systematics while maintaining a large field of view.  This design allows 
for full characterization of far-field optical performance using microwave sources on the ground.  
Here we describe the optical design of both instruments and 
report a full characterization of the optical performance and beams of \bicepp and the \keck at 150~GHz.
\end{abstract}


\section{Introduction} 

Inflation is a theory that describes the entire observable universe as a microscopic
volume that underwent violent, exponential expansion during the first
tiny fraction of a second (see~\cite{planckXXII} for a review).  Inflation is supported by the flatness and 
uniformity of the universe
observed through measurements of the cosmic microwave background (CMB).   
A generic prediction of inflation is the production of a gravitational-wave background, which 
in turn would leave a faint imprint in the polarization pattern of the CMB in addition to 
the already detected curl-free ``E-mode'' polarization sourced by density fluctuations at last scattering.  
A component of the inflationary signature would be a unique, divergence-free, ``B-mode'' polarization pattern
at large angular scales.
The strength of the B-mode polarization signature depends on the energy scale of inflation, and could be 
detectable if inflation occurred near the energy scale of grand unification, 
$\sim 10^{16}$~GeV~\citep{seljak97b,kamionkowski97,seljak97a}.

\bicepp and the \keck are microwave polarimeters that observe the CMB from the
South Pole in search of a B-mode polarization signature 
from inflation~\citep{ogburn10, sheehy10, instrument}. 
The receivers use a compact, on-axis refracting telescope 
design to couple radiation into a detector array of 512~antenna-coupled
transition edge sensor (TES) bolometers.  
\bicepp and the \keck leverage field-proven techniques employed for the \bicepo telescope~\citep{Keating03}, 
but with a vastly increased number of detectors, leading to increased sensitivity to the tiny 
B-mode polarization signal.  \bicepp has 512 detectors and the \keck has 2560 detectors in the
150~GHz-only configuration used in 2012 and 2013.

Table~\ref{table:receivers}
shows the changes in configuration for \bicepp and the \keck 
between observing seasons presented in this paper.
The \bicepp experiment 
observed for three years at 150~GHz from 2010 through 2012 and reported a detection of 
B-mode polarization on degree angular scales~\citep{results}.
All five \keck receivers were installed and operational beginning with the 2012 observing season.
For the 2012 and 2013 observing seasons, the \keck had five receivers at 150~GHz.  
Two of the \keck receivers were configured to observe at 95~GHz beginning with the 2014 observing season, 
and two more receivers have been configured
to observe at 220~GHz for the 2015 observing season.  \bicepp observes from
the Dark Sector Laboratory (DSL), and the \keck observes from a separate observing platform and telescope mount in 
the Martin A. Pomerantz Observatory (MAPO).  The two
observatories are situated approximately 200~m apart.  Here
we report on beam characterization for \bicepp and the 2012 and 2013 observing seasons of the \keck.  

This paper is one in a series of publications by the \bicepp and \keck collaborations. 
In this paper, we hereafter refer to other publications in this series as 
the Results Paper~\citep{results}, the Instrument Paper~\citep{instrument}, 
the Systematics Paper~\citep{systematics}, and the Detector 
Paper~\citep{detectors}. 

The Systematics Paper discusses in detail the level of contamination present in the \bicepp analysis 
presented in the Results Paper. The most significant systematic challenge arises from differential 
beam effects present in the \bicepp and \keck instruments.  Differential beam effects between 
co-located orthogonally polarized pairs of detectors
can lead to leakage of the CMB temperature signal into the much smaller polarization signal.  
It is therefore crucial that we fully understand the optical system and demonstrate that the achieved 
sensitivity is not compromised by systematics due to beam effects. 

In this paper, we describe in detail the optical design of the \bicepp and \keck telescopes (Section~\ref{sec:design}).
We report a characterization of the optical performance of 
the \bicepp and \keck instruments, compare it to physical optics simulations, and
discuss the level of E-mode to B-mode (E-to-B) leakage (Section~\ref{sec:characterization}).
We then discuss the construction of per-detector beam maps, which are inputs to simulations that are used to measure 
temperature to polarization leakage after removal of leading order contributions to beam mismatch between
co-located orthogonally polarized detectors in a pair (Section~\ref{sec:simulation}).

\begin{table*}
\begin{center}
\caption[\bicepp and the \keck receiver configuration]{\label{table:receivers}Configuration of the \bicepp 
  and \keck receivers for each year 
  of observation presented in this paper
 }
\begin{tabular}[c]{ccc}
\hline 
\multicolumn{2}{c}{Receiver} 
& Configuration\\
\hline \hline
\multicolumn{2}{c}{\bicepp 2010-2012}& 150 GHz, unchanged between observing seasons\\
\cline{1-3}
\multirow{6}{*}{The \keck 2012} & Receiver 0& 150 GHz\\
& Receiver 1&  150 GHz\\
& Receiver 2&  150 GHz\\
& Receiver 3&  150 GHz\\
& Receiver 4&  150 GHz\\
\cline{1-3}
\multirow{6}{*}{The \keck 2013} & Receiver 0 & 150 GHz, unchanged\\
& Receiver 1 & 150 GHz, one of the four focal plane tiles replaced\\
& Receiver 2 & 150 GHz, unchanged \\
& Receiver 3 & 150 GHz, focal plane from \bicepp installed\\
& Receiver 4 & 150 GHz, new focal plane installed\\
\hline
\end{tabular}
\end{center}
\end{table*} 

   \begin{figure}[ht]
   \begin{center}
   \begin{tabular}{c}
    \def\svgwidth{7cm}
    \input{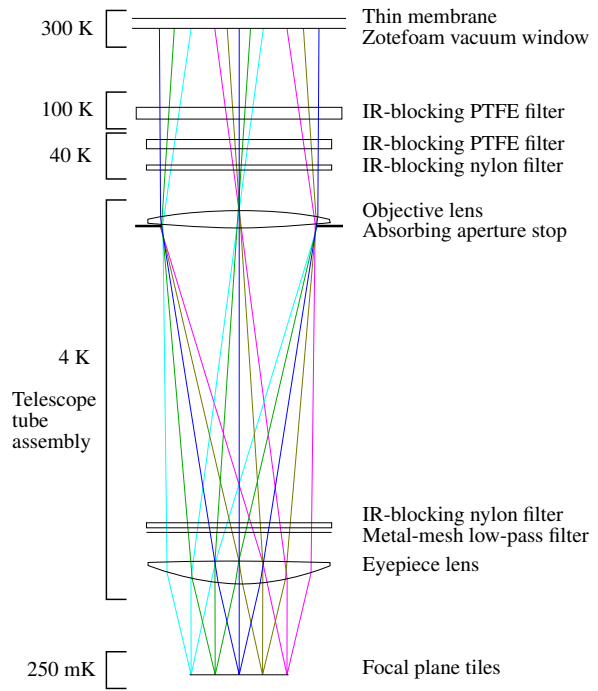}
   \end{tabular}
   \end{center}
   \caption{ \label{fig:opticsChain}
   A schematic of the \bicepp optical chain with each optical element labeled.  
   The relative position of the lenses and focal plane is to scale. The \keck optical chain is identical 
   except that the IR-blocking 
   filters in front of the objective lens are all on the 50~K stage in the \keck and the positions of the
   metal-mesh low pass filter and 4~K nylon filter are reversed.} 
   \end{figure}

\section{Optical design and modeling}
\label{sec:design}

The \bicepp/\keck optical design is a 
compact, single frequency band (150~GHz for \bicepp and 95, 150, or 220~GHz 
for the \keck), on-axis refractor with a 26.4~cm 
diameter aperture.
This design is similar to that of the \bicepo experiment~\citep{yoon06,takahashi10}.
The \bicepp/\keck optical design is schematically summarized in Figure~\ref{fig:opticsChain} and 
discussed in the Instrument Paper, \cite{aikin10}, and~\cite{vieregg12}.  

In both experiments, the aperture stop, lenses, and telescope tube assembly
are cooled to 4~K.  The \bicepp telescope is cooled with liquid helium to 4~K and 
has two vapor-cooled shields (at 40~K and 100~K respectively) as intermediate cryogenic stages, while the \keck 
telescope is cooled to 4~K using a pulse-tube cooler 
(Cryomech\footnotemark[1]\footnotetext[1]{http://www.cryomech.com} PT-410) 
and uses a single 50~K intermediate stage, 
also kept cold using the pulse-tube cooler.

The telescope tube is lined with microwave absorber 
(Eccosorb\footnotemark[2]\footnotetext[2]{http://www.eccosorb.com}~HR10) 
loaded with Stycast~2850 epoxy.  This design allows for stray, reflected light to be absorbed on the walls of the
telescope tube while providing minimal loading on the focal plane.  The lining of the 
telescope tubes has low reflectivity even at shallow incidence angles, and the epoxy-loading reduces the 
shedding of particulate matter from the absorber.

The absorbing aperture stop is made of microwave absorber (Eccosorb~AN74).
In the time reversed sense, pixels in the focal plane evenly illuminate the aperture
stop through the optics.  The central pixels terminate at $-$12~dB from their maximum, near their first Airy null
(for a description of beam shapes in the near field, see Section~\ref{sec:nearfield}).  
The 26.4~cm aperture provides $0.5^\circ$ FWHM diffraction-limited resolution on the sky, which we have chosen
in order to optimize 
the instrument to detect the peak of the primordial B-mode spectrum (at angular scales of 1--2$^\circ$).  

The compact design allows for full boresight rotation of the complete instrument.
In CMB observations, we observe at multiple boresight rotation angles, which provides cancellation
of a large class of systematic effects.  The ability to rotate the instrument around its 
boresight has proven to be a powerful tool to control the level of systematics contamination below the 
sensitivity of the experiments.

Section~\ref{sec:antenna} describes the antenna design, 
Section~\ref{sec:lens} describes the lens design, Section~\ref{sec:filter} describes the
design of the infrared (IR) filters, Section~\ref{sec:window} discusses the vacuum window, Section~\ref{sec:membrane}
describes the environmental membrane in front of the vacuum window, Section~\ref{sec:shield} discusses
the ground shielding system, and Section~\ref{sec:sim} describes the results of the optical simulation 
of the system.

\subsection{Antenna design}

\begin{figure}[ht]
   \begin{center}
   \begin{tabular}{c}
   \includegraphics[height=6cm]{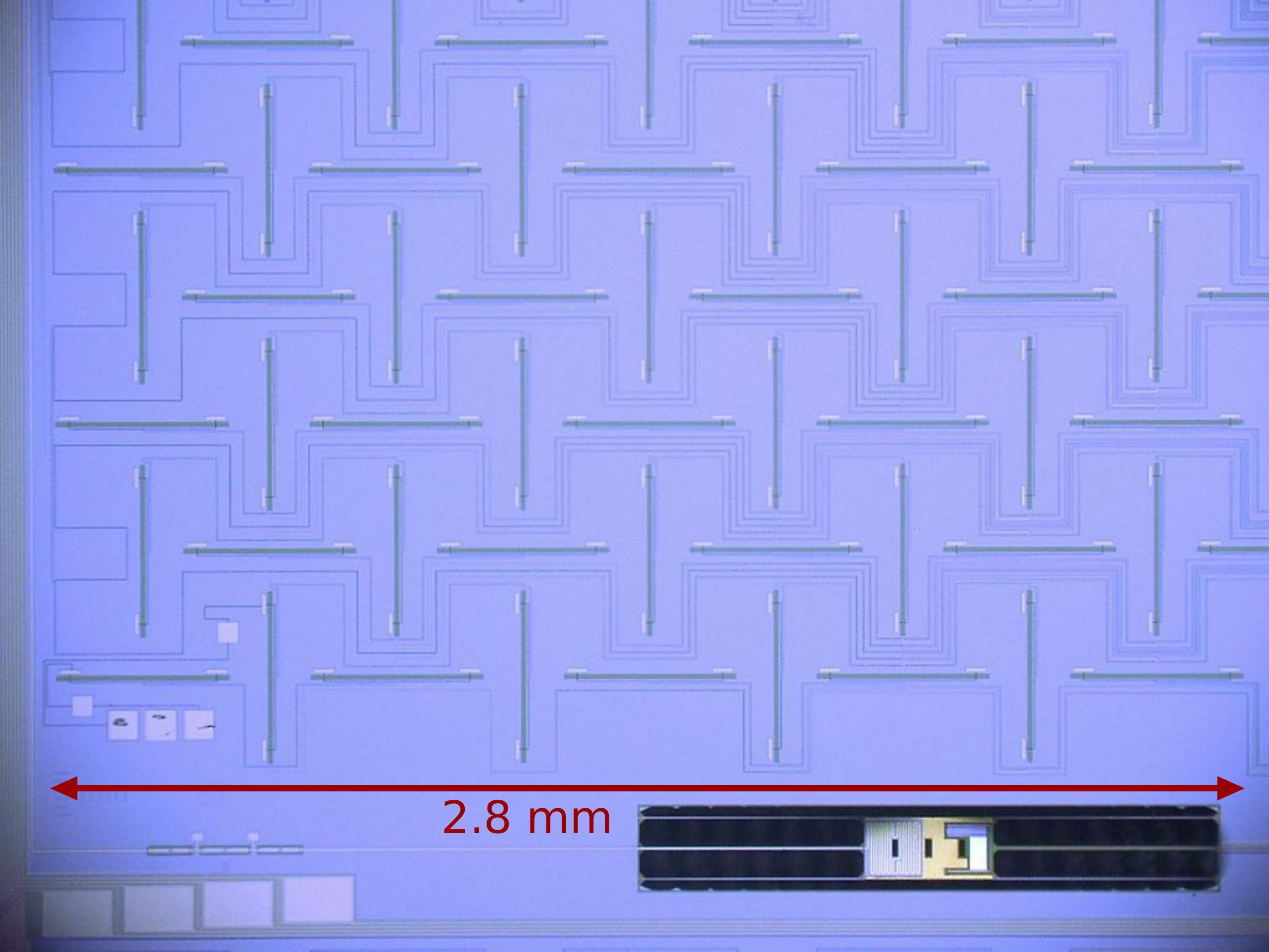}
   \end{tabular}
   \end{center}
   \caption{ \label{fig:antenna}
   A partial view of one \bicepp dual-polarization pixel, showing the band-defining filter (lower left), TES island (lower
   right), and part of the antenna network and summing tree.  For more information on the detector design, see the 
   Detector Paper.} 
   \end{figure}

Each pixel absorbs incident radiation through a planar superconducting antenna array, shown 
in Figure~\ref{fig:antenna}.
The antenna design used in \bicepp and the \keck is also
used in the balloon borne \spider experiment~\citep{spider},
and a similar design is used in 
\bicepq~\citep{bicep3}, which was deployed in late 2014 and will begin observation in early 2015.
The design is fabricated lithographically, allowing for rapid and scalable
production of detector tiles.
  
Microwave radiation is received by a two-dimensional array of sub-radiating slots that are spaced closely enough 
to avoid grating lobes at the high-frequency end 
of the observing band.  For 150~GHz detectors, this spacing is 604~$\mu\mathrm{m}$.  
We interconnect the slots with lithographed microstrip 
line and use the slot-array layer as a ground plane to 
shield those lines from direct stimulation by the incident radiation.  Each pixel is dual polarized, 
using orthogonally oriented but co-located sets of slots that function as independent antennas with 
independent feed networks.

We select the overall size of the antenna to match the optics in the telescope, placing the null between the 
main beam and the first Airy ring at the stop's edge.  For \bicepp and the \keck's 150~GHz cameras,
the antenna array is 7.8~mm on a side, using a $12\times12$ array of slots in each polarization.

By adjusting the impedance of the microstrip lines in the feed network, we could generate an arbitrary 
illumination pattern and control sidelobe levels.  However, for simplicity of design in \bicepp and the \keck, 
we feed 
power from slots uniformly in the feed network.  This top-hat illumination creates $<-$12~dB sidelobes that 
terminate on the absorbing stop.  The feed combines waves from the slots with uniform phase 
to create circular Gaussian beams with matching centroids between detector pairs.

After summing radiation in the feed network, power passes through the integrated band-defining filter 
before terminating in a lossy transmission line on a thermally isolated island also 
occupied by a TES bolometer. For more information on the detector design, see the Detector Paper.

\label{sec:antenna}

\subsection{Lens design}
\label{sec:lens}
The refracting optics consist of a pair of lenses that are cooled to 4~K and are made of high-density polyethylene 
(HDPE), a material with low loss and an index of refraction of 1.54 at millimeter wavelengths~\citep{lamb}.  
The aperture stop is 
coincident with the objective lens, which sits at the focus of the eyepiece lens, 
defining a telecentric system that makes the plate scale 
robust against any as-built defocus of the telescope.  We chose to have only two lenses
to minimize partial reflections and ghosting.

To minimize the radius of curvature of the eyepiece lens for ease of fabrication, the 
eyepiece lens (26.0~cm diameter) is located as far from the focal plane (18.0~cm diameter) 
as it can be without vignetting beams from any detectors in the focal plane. The distance to 
the eyepiece lens was set to 15.0~cm.

We chose the lens separation, and thus the focal length, to be 55.0~cm.  This directly produces a 
plate scale such that the angular resolution of the telescope on the sky corresponds roughly to 
twice the physical separation between pixels on the focal plane, allowing us to Nyquist sample modes on the sky.  
Physical optics simulations performed with
Zemax\footnotemark[3]\footnotetext[3]{http://www.zemax.com} optical design
software showed that shorter focal lengths introduced aberrations for corner detectors on the focal plane 
and that longer focal lengths 
resulted in asymmetric illumination of the aperture 
stop that corresponded to elliptical far-field beam patterns.
  
Using time-forward optics simulations with 
collimated ray bundles distributed in the same way as the detectors across the focal plane and incident 
on the objective lens at angles from $0^\circ$ to $8^\circ$ from normal incidence, 
we selected the objective lens geometry that minimizes 
the beam waist at the eyepiece lens.  Analogously, we used time-reverse simulations of Gaussian-profile weighted 
collimated rays incident on the eyepiece lens at angles of $0^\circ$ to $10^\circ$ from normal incidence to 
choose the eyepiece lens 
geometry that minimizes the beam waist on the objective lens.  We solved these iteratively in Zemax to converge on 
acceptable lens surface geometries.  We found that while perfectly telecentric systems had good image quality,
such designs illuminate the aperture asymmetrically and can induce far-field ellipticity.  
We ultimately sacrificed telecentricity to attain symmetric far-field beams.

We anti-reflection (AR) coated the \bicepp/\keck lenses with porous 
Teflon (Mupor\footnotemark[4]\footnotetext[4]{http://www.porex.com})
whose index of refraction and thickness were customized to tune the performance to the 
observing frequency of each receiver given the optical material being AR coated.  The AR coating was
attached to each optical element through heat bonding using a thin film of low-density polyethylene (LDPE).

   \begin{figure*}[ht]
   \begin{center}
   \begin{tabular}{c}
   \includegraphics[height=5cm]{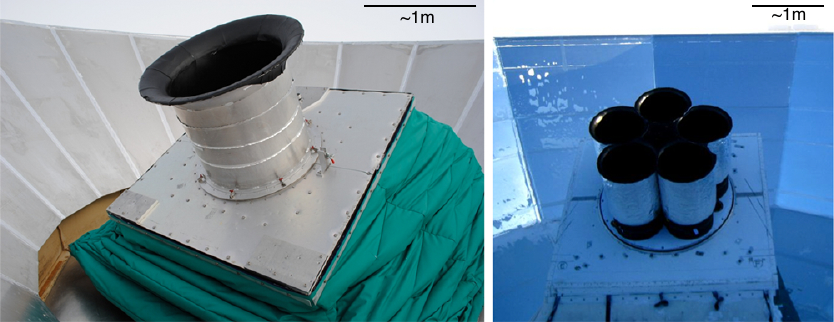}
   \end{tabular}
   \end{center}
   \caption{ \label{fig:picture}
   A picture of \bicepp (left) and the \keck (right) from the outside.  The forebaffles and the reflective ground shield 
   are visible.} 
   \end{figure*}
   \begin{figure*}[ht]
   \begin{center}
   \begin{tabular}{c}
   \includegraphics[width=16cm]{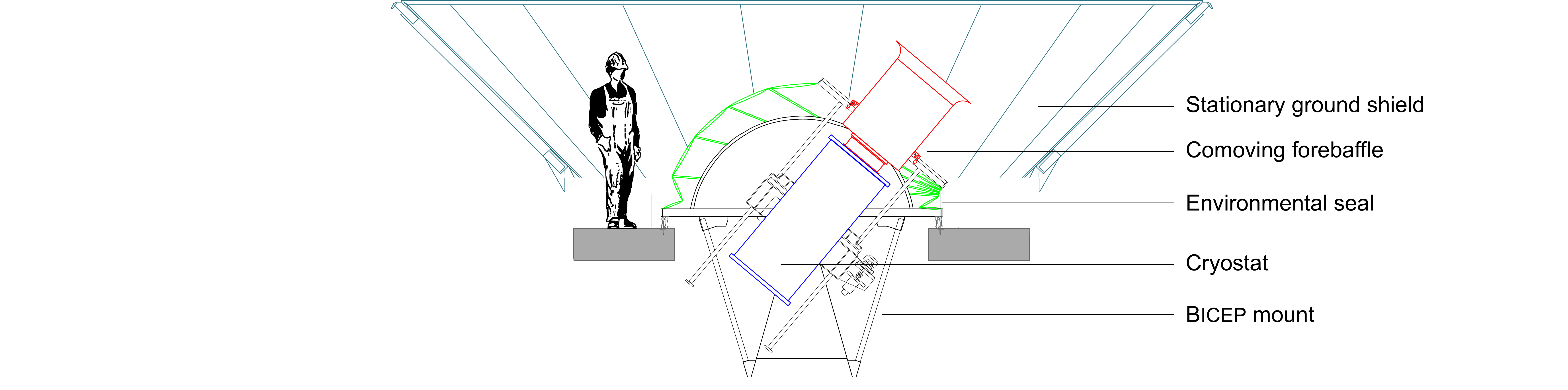} \\
   \includegraphics[width=16cm]{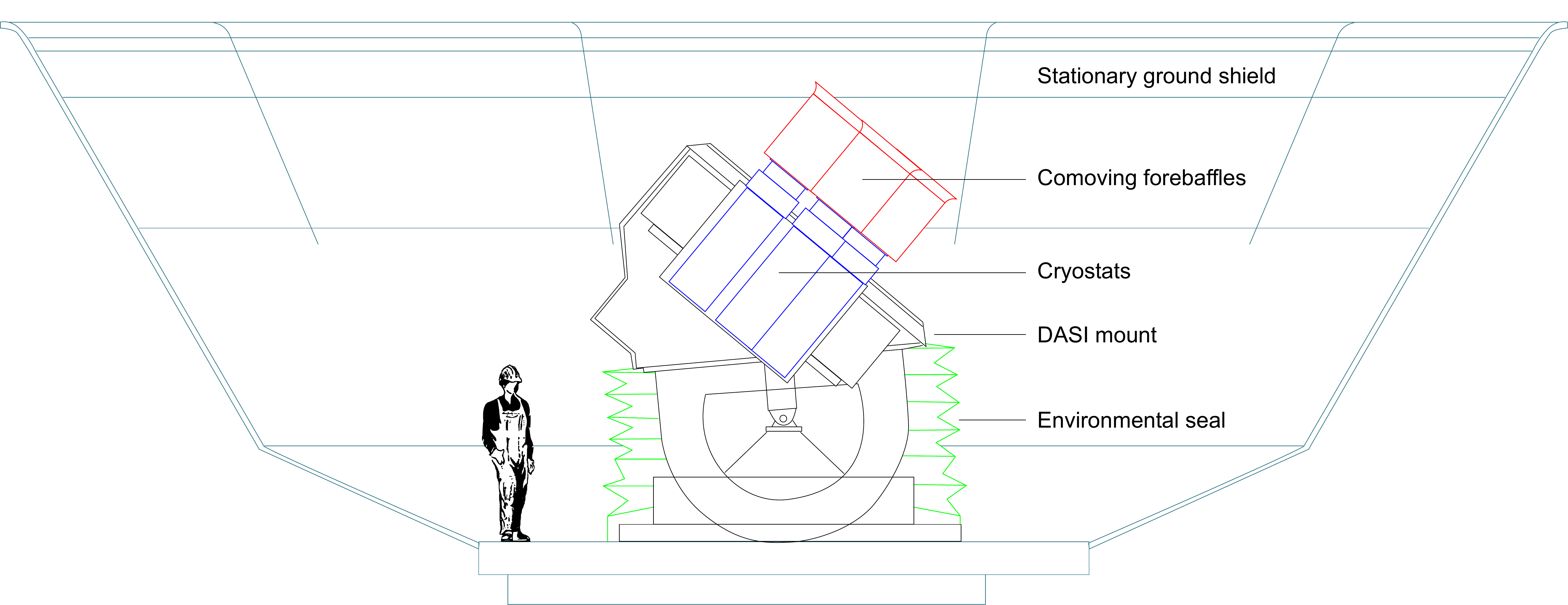}
   \end{tabular}
   \end{center}
   \caption{ \label{fig:combo_mounts}
   Cross-sectional view of \bicepp (top) and the \keck (bottom) in their observing configurations.
   Both experiments are shown to the same scale.
   The receivers are shown in blue, the forebaffles are shown in red, and the environmental seals are shown in
   green.  Also shown are the mounts and the stationary ground shields. }
   \end{figure*}

\subsection{Filter stack}
\label{sec:filter}
Teflon, nylon, and a metal-mesh low-pass filter block IR radiation 
from reaching the focal plane.  Teflon has excellent in-band transmission at cryogenic 
temperatures.  While nylon has higher in-band transmission 
loss, it has a steeper transmission rolloff out of band, 
providing significant reduction of far-IR loading on the 
sub-kelvin stages.  The metal-mesh filter is a low-pass filter with a cutoff at 250~GHz, providing 
additional blocking of out-of-band power~\citep{ade06}.   

In \bicepp and the \keck, two Teflon filters and one nylon filter sit in front of the objective lens.  
In \bicepp, one Teflon filter is held at 100~K and 
the second Teflon filter and the nylon filter are held at 40~K, while in
the \keck they are all on the 50~K stage.  In both experiments, 
a nylon filter and the low-pass 
metal-mesh filter sit at 4~K between the objective and eyepiece lenses 
inside the telescope itself to further reduce loading.
We AR coat all filters using the same process described for AR coating the lenses in Section~\ref{sec:lens}.

\subsection{Vacuum window}
\label{sec:window}
The \bicepp/\keck vacuum window 
has a 32~cm clear aperture, making design and construction of strong and durable 
windows a challenge.  The \bicepp vacuum window was made of 
Zotefoam\footnotemark[5]\footnotetext[5]{http://www.zotefoams.com}~PPA30, 
a dry nitrogen-expanded polypropylene 
foam chosen for its high microwave transmission, its strength against deflection under vacuum, its 
adhesion strength to the epoxy used to bond the foam to an aluminum frame, and
its previous successful use in similar applications~\citep{runyan2003,takahashi10}.
The 10~cm thick window was made of four layers of PPA30 joined together by heat lamination and
was bonded to an aluminum frame using Stycast 1266 epoxy.
We measured the transmission of 
the \bicepp window to exceed 98\% at 150~GHz, consistent with the \bicepo 
window~\citep{takahashi10}. 

\keck windows are made of
Zotefoam HD30 because Zotefoam ceased production of~PPA30.  HD30 is a nitrogen-expanded
polyethylene foam that has similar performance to PPA30 in both microwave transparency 
and mechanical strength, with slightly inferior lamination and adhesion qualities.
\keck windows are composed of layers of HD30 joined together by heat lamination, like
the \bicepp window.
The vacuum windows for the 2012 and 2013 observing seasons were bonded to their aluminum frames
using Stycast 2850 epoxy.
After the 2012 observing season, we increased the thickness of the windows to 12~cm 
(the maximum possible with a four layer HD30 laminate)
because we observed the vacuum window foam 
tearing away from the epoxy used to attach the foam to the aluminum frame.  
The $\lesssim 2\% $ transmission loss appears to be dominated by the laminations, so
maintaining four layers while increasing the thickness has had minimal impact on performance, 
while the thicker windows are able to 
withstand the force from the vacuum over time.

\subsection{Membrane}
\label{sec:membrane}
In front of the window sits a thin transparent membrane held tautly by two aluminum rings.  The membrane
creates an environmental shield to protect the window from snow.  The enclosed space between the membrane 
and the window is slightly pressurized with dry nitrogen gas to prevent condensation on the foam window.
Room-temperature air flows through holes in the aluminum ring onto the top of the membrane so that any snow
that accumulates sublimates away.

For \bicepp, the initially deployed membrane was 0.5~mil thick Mylar, which has low reflectivity at
150~GHz (0.2\%).  In April 2011, the membrane was replaced with 0.9~mil biaxially oriented polypropylene
(BOPP) and the pressure of the nitrogen was adjusted to reduce vibrations in the membrane, discussed in the
Instrument Paper.  
The \keck uses the same BOPP membranes as \bicepp.

\subsection{Baffling and ground shielding}
\label{sec:shield}
A co-moving ground shield, called the forebaffle, installed in front of each receiver's 
vacuum window reduces sidelobe pickup.  The forebaffles for \bicepp and the \keck have identical
construction except for their overall sizes. The \bicepp forebaffle was 94~cm tall and 71~cm
in diameter, and the \keck forebaffles used for the five-receiver configuration (2012 and later)
are 74~cm tall and 64~cm in diameter.  The forebaffles have rolled lips at the upper edge of the cylinder 
to reduce diffraction and diffuse any far-sidelobe beams.  
We coat the inside of the forebaffle with Eccosorb HR10 microwave absorber and Volara, a 
weatherproofing foam, to terminate the 
sidelobes while only modestly increasing thermal loading on the focal plane.  The forebaffles intersect radiation 
at $9.5^{\circ}$ from the telescope boresight axis as measured from the edge of the vacuum window.  
The forebaffles and ground shield for \bicepp and the \keck are shown in the photographs
in Figure~\ref{fig:picture} and in cross-sectional diagrams in Figure~\ref{fig:combo_mounts}.

A fixed reflective ground shield, visible in Figure~\ref{fig:picture}, redirects any stray light to the cold
sky and shields the telescope from having any direct line-of-sight from the ground.  
The \bicepp ground shield was previously used
by \bicepo~\citep{takahashi10} 
and the \keck ground shield was previously used by \dasi~\citep{dasi} and~\quada~\citep{quad}. 

The two ground-shielding stages were designed so that at the lowest CMB 
observation angle (an elevation of $55^\circ$), the top of the fixed reflective 
ground shield is still $11^\circ$ below any direct line of sight from the telescope past the 
co-moving forebaffle.  Additionally, as viewed from the receivers, the top of the fixed reflective 
ground shield is at least $15^\circ$ above the ground.
Therefore, rays from the telescope must diffract twice (over
the edge of the co-moving forebaffle and the ground shield) before they hit the ground. 
A drawing of \bicepp and \keck receivers in their observing configurations is 
shown in Figure~\ref{fig:combo_mounts}.

\subsection{Modeled far-field beams}
\label{sec:sim}
\begin{figure}[ht]
   \begin{center}
   \begin{tabular}{c}
   \includegraphics[height=5.3cm]{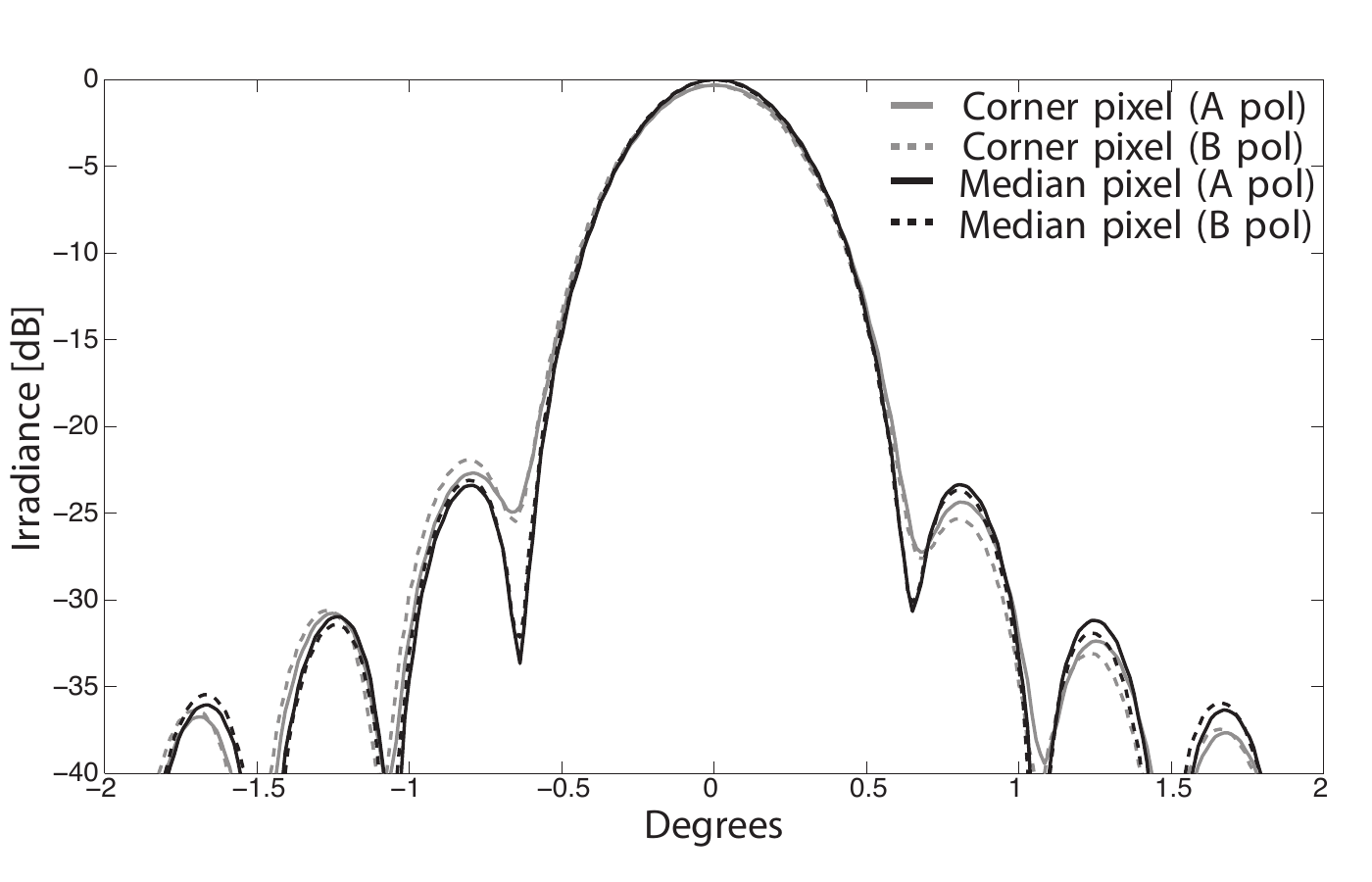}
   \end{tabular}
   \end{center}
   \caption{ \label{fig:zemax}
   Simulated far-field beam pattern for orthogonally polarized $A$ and $B$ beams, for both median-radius and 
   corner pixels in the focal plane.  Beams share the same peak normalization.} 
   \end{figure}

We have modeled the far-field beam patterns with Zemax simulations that account for optical elements 
between the focal plane and the aperture stop, including lenses and filters.
The optical simulation is discussed more fully in~\cite{aikin10}.  Figure \ref{fig:zemax} shows the simulated
far-field beam pattern for the two orthogonally polarized beams in a given
detector pair, which we denote ``$A$'' and ``$B$.''  The Figure shows median-radius and corner 
pixels in the focal plane 
to demonstrate beam uniformity across the focal plane.  The median-radius pixel is displaced 5.6~cm from 
the optical axis, and the corner pixel is displaced 8.0~cm.  Beams are Gaussian with a $0.5^{\circ}$ FWHM,
and the Airy ring structure is clearly visible.

\section{Optical characterization}
\label{sec:characterization}
We have fully characterized the beams of \bicepp and the \keck for its 2012 and 2013 configurations.  The ultimate
goal of this characterization is to place constraints on the temperature to polarization leakage and the 
E-to-B leakage due to beam effects, as presented in the Systematics Paper.  

In this Section,
we first present results of near-field beam characterization studies (Section~\ref{sec:nearfield}).  We then discuss
the far-field beam characterization campaign at the South Pole (Section~\ref{sec:beamchar}), including measurements of beam
shape parameters (Section~\ref{sec:beamshape}) and differential beam parameters (Section~\ref{sec:diffbeam}) and 
a comparison with optical models (Section~\ref{sec:simCompare}).  Finally, we discuss power contained in far sidelobes
(Section~\ref{sec:farsidelobes}) and polarization angle and cross-polar response 
measurements (Section~\ref{sec:polangle}), 
including placing constraints on E-to-B leakage.

\subsection{Near-field beam characterization}
\label{sec:nearfield}
To characterize aperture illumination, we measured the near-field beam pattern of each \bicepp and \keck
detector. While far-field
beam maps are primarily sensitive to the amplitude distribution of the electric field in the focal
plane, maps made in the aperture plane of the instrument are primarily sensitive to the phase of the
electric field in the focal plane. As a result, near-field maps can serve as a probe of phase gradients
within the phased-array antennas. Near-field maps can also serve as a probe of secondary reflections
that focus near the aperture plane and of vignetting within the telescope.  The main purpose of
near field measurements is to feed back into the detector and optics fabrication process for future generations
of receivers.

Near-field maps were made using a chopped thermal source mounted on an x-y translation stage 
attached to the cryostat above the vacuum window as close to
the aperture stop of the telescope as possible.  In practice, the 
source is about 30~cm above the aperture.  

Near-field maps were acquired during two consecutive summer seasons at the South Pole for \bicepp
and are acquired after the first cool down of each \keck receiver at South Pole.
The TES detectors can operate on each of two superconducting transitions: 
the titanium transition, on which CMB observations are made, and the higher-temperature
aluminum transition.
Beam mapping data is acquired with the TES detectors on their
aluminum transition, which can accommodate the higher loading present in the lab.  
We observe the CMB with the detectors on their titanium transition, but the beams formed for a given
channel are expected to be the same for the two superconducting transitions, since beams are formed by the 
antenna network and the optics, not the detector itself.

\begin{figure}[ht]
   \begin{center}
   \begin{tabular}{c}
   \includegraphics[height=6.7cm]{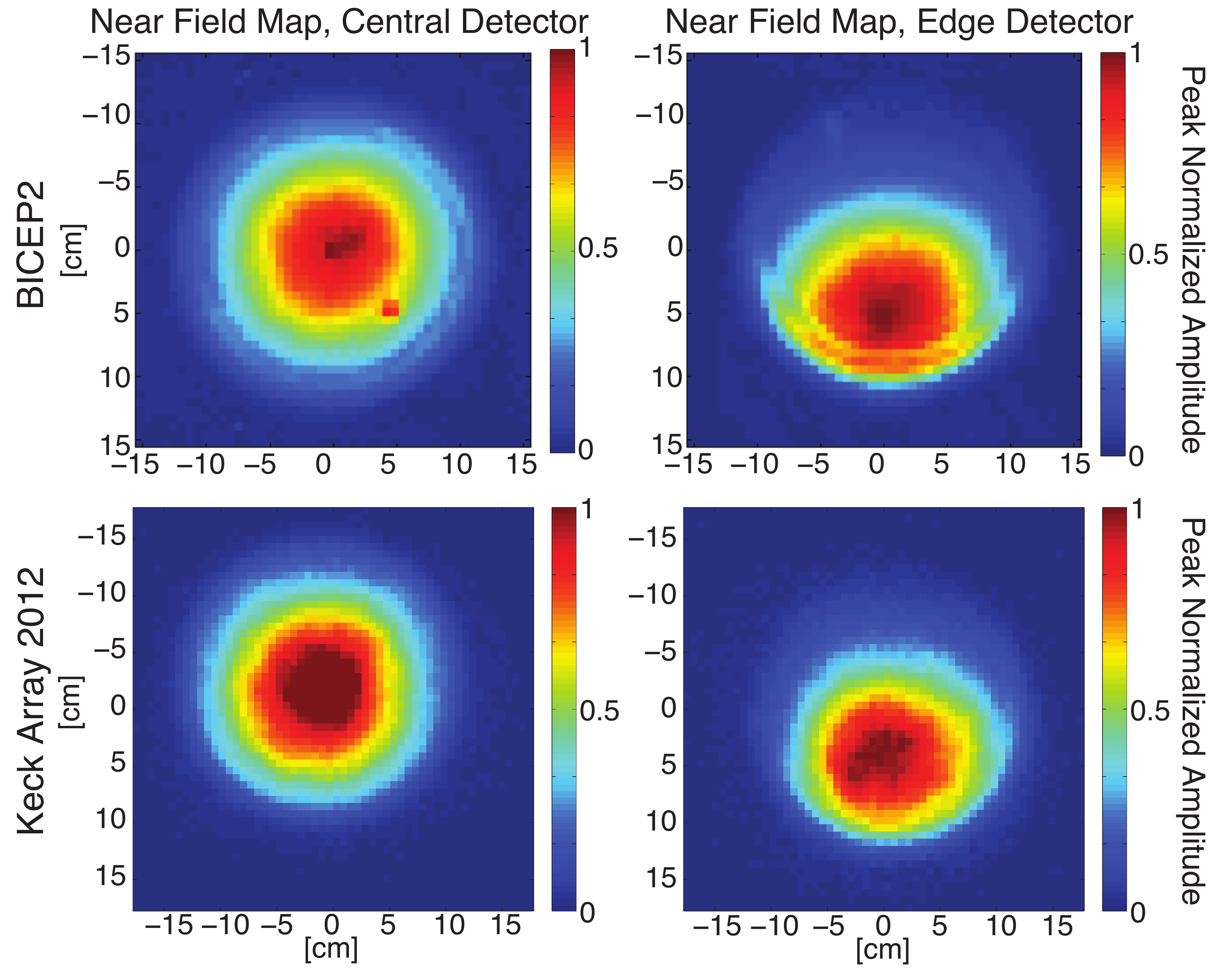}
   \end{tabular}
   \end{center}
   \caption{ \label{fig:nearField} A measurement of the near-field beam pattern for two example channels
     in \bicepp and the \keck during 2012.  Left: A detector near the center of a tile in the
     focal plane. Right: A detector near the edge of the
     focal plane, showing significant truncation by the aperture (worst case).  This truncation leads to
     moderate ellipticity in the far field.
     The sharp feature seen for the center pixel in \bicepp is consistent with a
     reflection off of the 4~K nylon filter, and is diffusely coupled to the sky.
   } 
   \end{figure}

Figure~\ref{fig:nearField} shows
the beam pattern of two example detectors
in the near field for \bicepp and the \keck in its 2012 configuration.  
In each case, the left panel shows the 
beam pattern of a detector near the center of a tile in the focal plane that evenly 
illuminates the aperture.  The right panel shows the beam pattern of a detector 
near the edge of the focal 
plane that is significantly truncated by the aperture because of non-ideal 
beam pointing at the focal plane, which we call ``beam steer.''  This beam steer
can be as large as 5--10$^\circ$ in \bicepp, 
more than is predicted by the physical optics model presented
in Section~\ref{sec:sim}~\citep{aikinThesis}.
This impacts not only
optical throughput, but can also potentially introduce beam distortion  
caused by the asymmetric and aggressive illumination of the aperture stop. 
This type of truncation translates to moderate ellipticity in
the beam pattern in the far field and only affects a small fraction of detectors that sit 
near the edges of tiles in the focal plane.
Beam steer
is typically smaller for \keck receivers compared to \bicepp.  

As discussed in Section~\ref{sec:simulation}, systematic effects stemming from  
the observed per-beam far field ellipticity have been
successfully removed in analysis to the level required
for \bicepp and the \keck.  Although the effects of beam steer are therefore not concerning
for \bicepp and the \keck, reducing beam steer for future generations of detectors would 
increase optical throughput and reduce far field ellipticity, improving sensitivity slightly
and potentially improving the achievable level of residual systematics.

   \begin{figure}[ht]
   \begin{center}
   \begin{tabular}{c}
   \includegraphics[width=8.4cm]{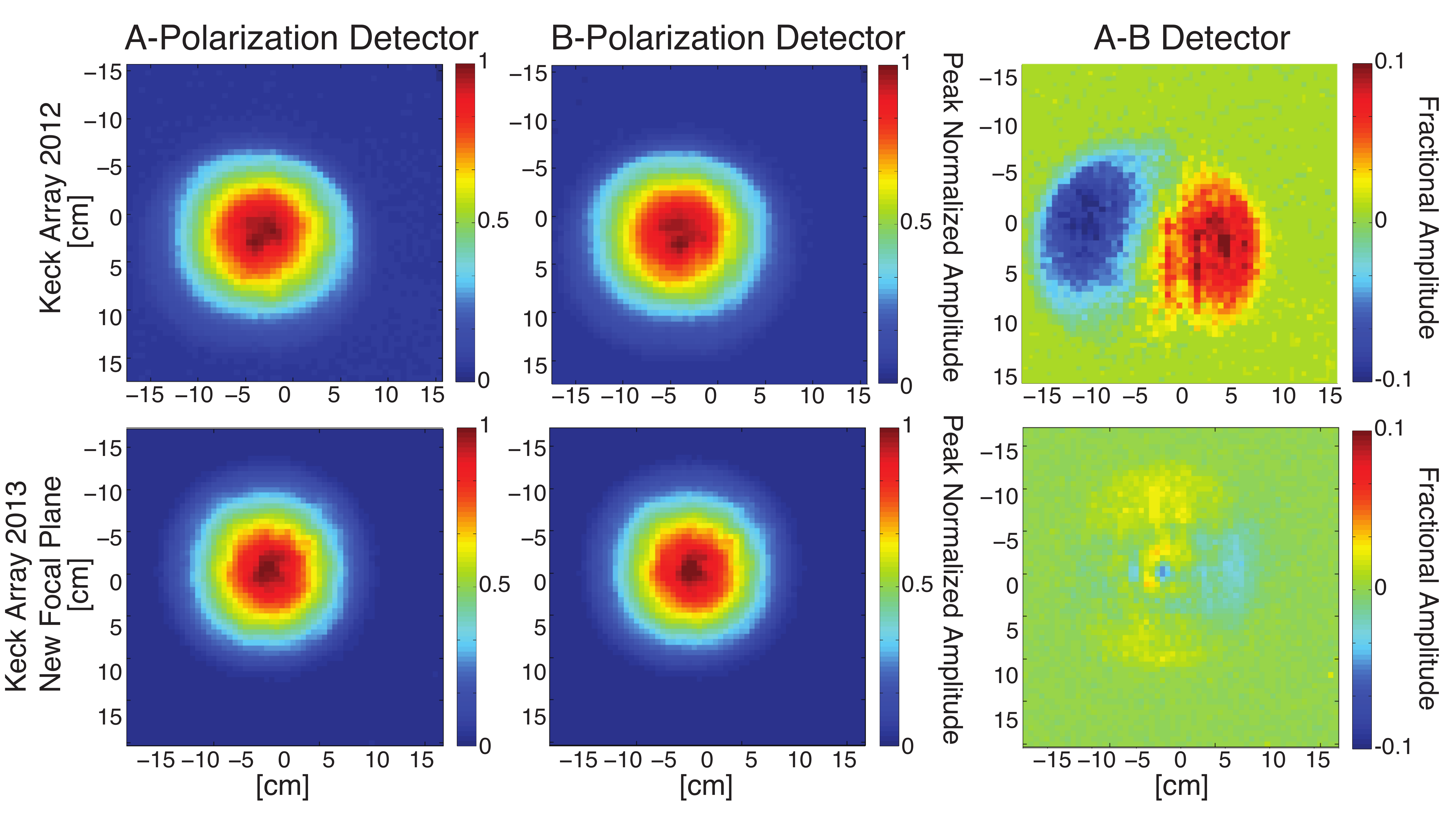}
   \end{tabular}
   \end{center}
   \caption{ \label{fig:nearFieldDiffKeck}
     An example \keck detector pair that shows beam mismatch in the near field. Left: The optical response of
     an $A$ polarization detector in a typical detector pair.  
     Center: The optical response of the co-located $B$ polarization
     detector.  Right: The fractional difference between the $A$ and $B$ optical response. The top panels show a typical
     detector pair from a focal plane
     in 2012, and the bottom panels show a typical detector pair from the new focal plane installed in 2013 
     with dramatically reduced
     differential pointing.} 
   \end{figure}
   \begin{figure*}[ht]
   \begin{center}
   \begin{tabular}{c}
   \includegraphics[height=4.5cm]{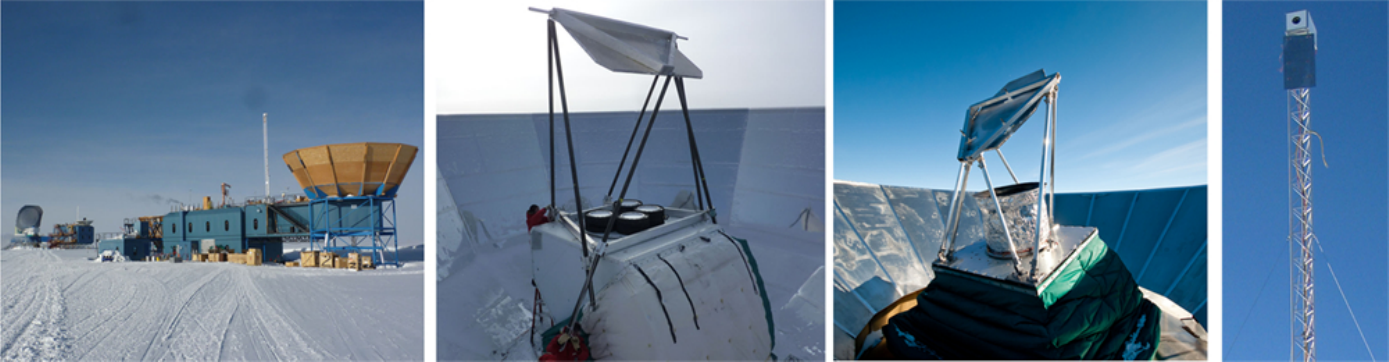}
   \end{tabular}
   \end{center}
   \caption{ \label{fig:farFieldMapping} The setup for measuring far-field beam patterns of \bicepp and \keck 
     detectors {\it in situ} 
     at the South Pole.  A chopped thermal source broadcasts from a mast on MAPO or DSL, 
     and a large aluminum honeycomb mirror
     is installed to redirect the beams of \bicepp and the \keck to the source. Left: DSL (far) and MAPO (near),
     home of \bicepp and the \keck.  Both masts are up and are being used for beam mapping.  Center Left:
     The large mirror installed for beam mapping on the \keck. Center Right: The smaller
     mirror installed for beam mapping
     on \bicepp. Right: A close up of a microwave source mounted on the mast.
   } 
   \end{figure*}

The sharp bright
feature in the bottom right quadrant of the \bicepp central detector map is consistent with a 
secondary reflection from the 4~K
filters refocused into the aperture plane. This spot contains less than 0.1\% of the integrated power
of the main beam.    Moreover, since it forms a sharp focus in the aperture plane, it must be broadly
and diffusely coupled to the sky in the far field. This feature is not present in \keck receivers.

Figure~\ref{fig:nearFieldDiffKeck} shows near-field 
maps for a typical orthogonally polarized co-located detector pair
in the \keck, as well as a difference map between the two detectors in the pair.  
The top panels show a typical pair of detectors
from a \keck focal plane in 2012, and the bottom panels show a typical pair from a new focal plane installed in 2013.
Beams measured in the near field in \bicepp and early \keck focal planes 
show a consistent mismatch in the $A$ and $B$
beam centroids of co-located, orthogonally polarized detector pairs. 
The centroid displacement is consistently co-aligned
with the polarization axes of each tile, and thus also the summing tree axes. The amplitude was
measured to be nearly constant across the focal plane, except for a small subset of pixels suffering from the
severe beam steer illustrated in Figure~\ref{fig:nearField}.  The observed patterns are consistent with 
our detector modeling, described in~\cite{roger}.

Mismatch in the near-field centroids alone will not
lead to any substantial far-field beam mismatch. While the beams may be displaced in the near field,
the resulting angular displacement on the sky is negligible. 
However, non-idealities
in the optics of the instrument, such as birefringence in the optics or an out-of-focus system,
can serve to translate a near-field mismatch to the far field.

Detector development efforts have greatly reduced the near-field mismatch. 
The component of mismatch parallel to the summing trees was reduced by 
increasing spacing between lines of the summing tree to reduce the parasitic 
coupling and resulting phase errors.  We have reduced the remaining phase error 
from residual coupling by adding phase lags to the summing tree.  The additional 
path length corrects the residual phase step across the antenna's mid-plane.  
Subsequent detector testing has shown that the source of the remaining near-field 
differential pointing, predominantly along the axis orthogonal to the summing trees, 
is related to contamination in the niobium films of the microstrip lines, producing 
non-constant wave speeds across the feed.  The efforts to improve
the matching of phased-array antenna beams in the aperture plane are described in detail in the 
Detector Paper.  \keck focal planes installed in late 2012 and later
have dramatically reduced near-field mismatch beginning with the 2013 observing season 
as a result of these efforts (see the lower panels of Figure~\ref{fig:nearFieldDiffKeck}).

\subsection{Far-field beam characterization}
\label{sec:beamchar}
   \begin{figure*}[ht]
   \begin{center}
   \begin{tabular}{c}
   \includegraphics[height=7cm]{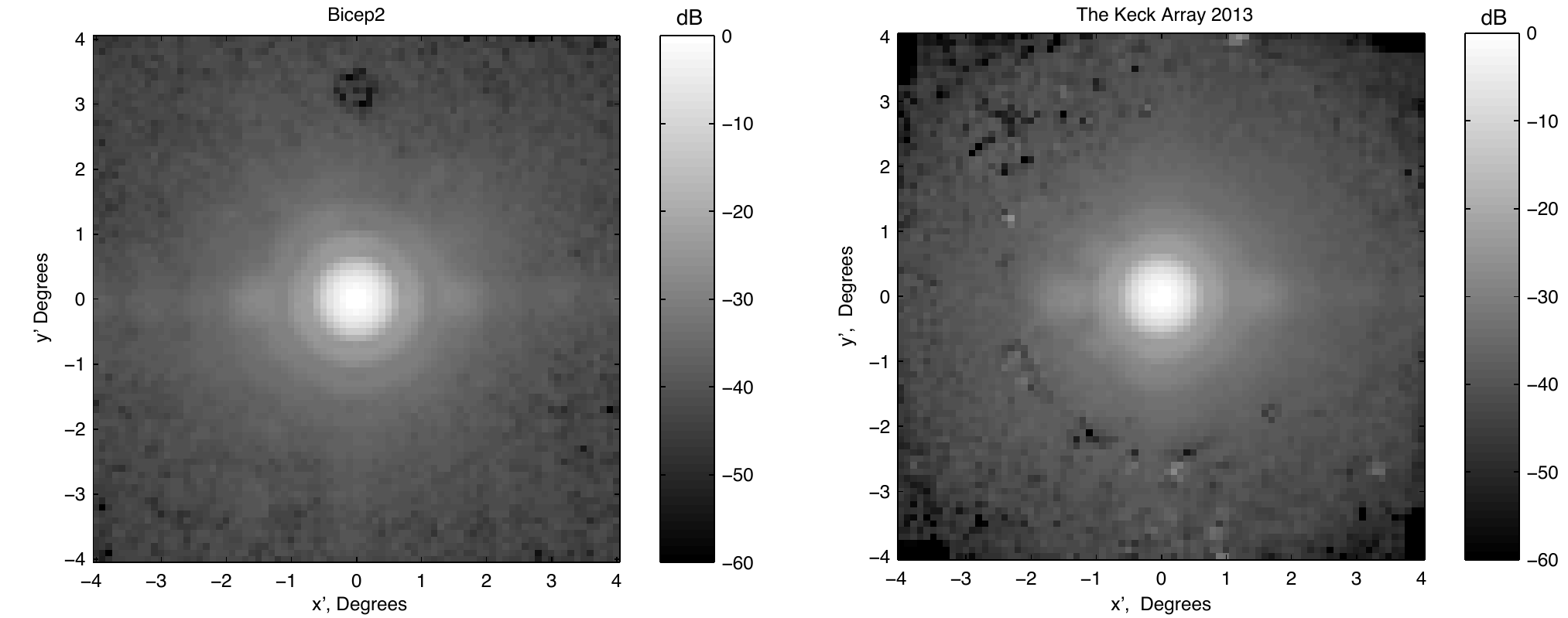}
   \end{tabular}
   \end{center}
   \caption{ \label{fig:stackedBeams}  Left: A map of the \bicepp far-field response made with a thermal 
     source, centered, rotated, 
     and co-added over all operational channels from 2012 data.  Right: A map of 
     the \keck far-field response made with a 
     thermal source, centered, rotated,
     and co-added over all operational channels for all 2013 \keck receivers.
     The color scale is logarithmic with decades marked in dB.
     The measured main beam shape and Airy ring structure are well-matched by simulations 
     (see Section~\ref{sec:simCompare}).  Cross-talk features 
     are evident, $1.5^\circ$ to the right and left of the main beam.
     The feature $3^\circ$ above the main beam in the \bicepp is also due to crosstalk.
   } 
   \end{figure*}
   \begin{figure}[ht]
   \begin{center}
   \begin{tabular}{c}
   \includegraphics[height=4.8cm]{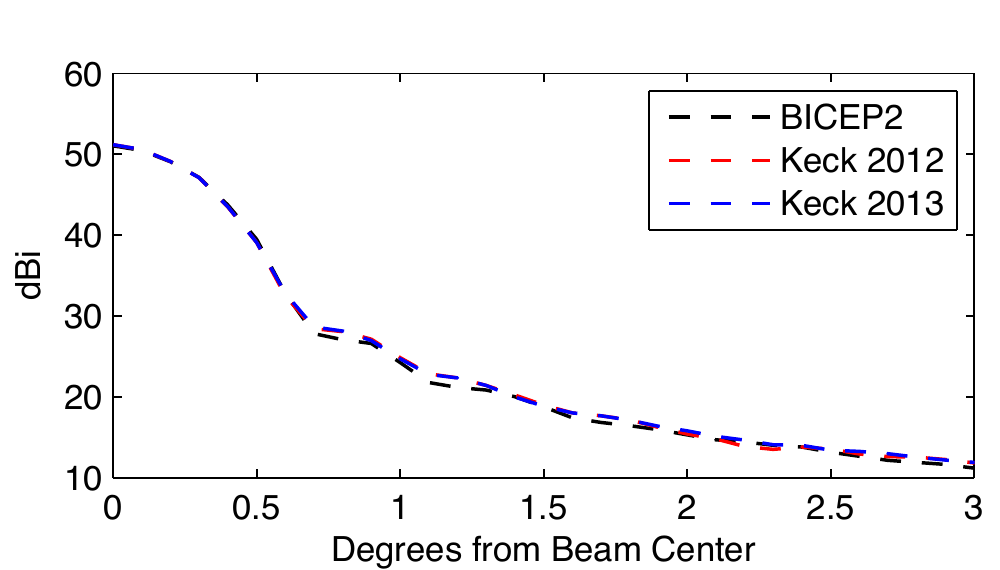}
   \end{tabular}
   \end{center}
   \caption{ \label{fig:beamProfileKeck} Azimuthally averaged beam profiles for \bicepp and
     the \keck for 2012 and 2013, co-added over all operational channels.
   } 
   \end{figure}
We are able to fully characterize the beam of each detector in the far field using
microwave sources on the ground because the far field of the telescopes is only at a distance of
70~m.  We define the far field to be a distance of $2D^2/\lambda$, where
$D$ is the size of the aperture (26.4~cm) and $\lambda$ is the wavelength at the observing
frequency of 150~GHz (2~mm).

We measure the optical response
through an extensive beam mapping campaign at the South Pole.  Figure~\ref{fig:farFieldMapping}
shows the setups used to measure the beam pattern in the far field.  For beam mapping, we
install flat aluminum
honeycomb mirrors to redirect the beams over the top of the ground shields and to an unpolarized chopped thermal
source mounted on a 10~m tall mast on MAPO
(195~m away for \bicepp) or DSL (211~m away for the \keck). 
We used both a small aperture source (20~cm) and a large aperture source (45~cm) for beam characterization.
These sources appear as point-like in the far field;
the large aperture source subtends an angle of $0.1^\circ$ as viewed by the telescope, which is much smaller
than the size of the beam.
The thermal sources chop between a 
mirror directed to zenith ($\sim15$~K) and ambient temperature ($\sim250$~K) at a tunable frequency,
set to be 18~Hz for the smaller aperture source and 10~Hz for the larger aperture source. 
We use maps made with these sources
to study the beam shapes of each detector in order to fully understand the 
effects of the shape of the beams on the science data.
We also made beam maps using the broad-spectrum noise source described
in Section~\ref{sec:rps},
which has a large and adjustable amplitude and is good for measuring relatively dim 
sidelobe features.

Far-field maps are made by rastering in azimuth at a scan rate of $2.0^\circ$ per second 
and stepping in elevation at a fixed boresight rotation angle.
We make maps at multiple boresight rotation angles to check that a
rotation of the receiver does not affect the results of the measurement.  
Because the effective position of the source is only $\sim2^\circ$ above the horizon, and we mask out
the ground when making maps,
multiple boresight angles allow more thorough mapping of the response at large 
angles from the center of the beam.
All measurements are
made with the detectors on the aluminum transition. 
We take beam map data at a high sampling rate (150~Hz), filter the chop reference signal to match the filtering
that occurs in the readout system, and then demodulate the timestream data. The reflection off of the mirror and
parallax effects are handled with a pointing model that describes the \bicepp and \keck mount systems as well as the
mirrors used for beam mapping~\citep{aikinThesis}. 

Figure~\ref{fig:stackedBeams} shows maps of the far-field response 
of \bicepp and the \keck in its 2013 configuration, 
centered, rotated, and co-added over all operational channels.
The maps have been rotated before co-addition to
account for the boresight angle rotation and the 
clocking of \keck receivers in the drum on the mount
so that co-added maps are fixed to focal plane coordinates.
The measured main beam shape and Airy ring structure are well-matched by simulations 
(see Section~\ref{sec:simCompare}),
and cross-talk features are evident in the stacked maps ($1.5^\circ$ from the main beam), corresponding
to the location of the nearest neighbor detector in the multiplexing readout scheme.
Figure~\ref{fig:beamProfileKeck} shows the azimuthally averaged beam profile for \bicepp and the
\keck in its 2012 and 2013 configurations, derived from the co-added maps.

From these measurements, we extract beam shape parameters
on a per-detector basis and characterize the difference in the beams between detectors in each
detector pair.

   \begin{figure*}[ht]
   \begin{center}
   \begin{tabular}{c}
   \includegraphics[height=4.8cm]{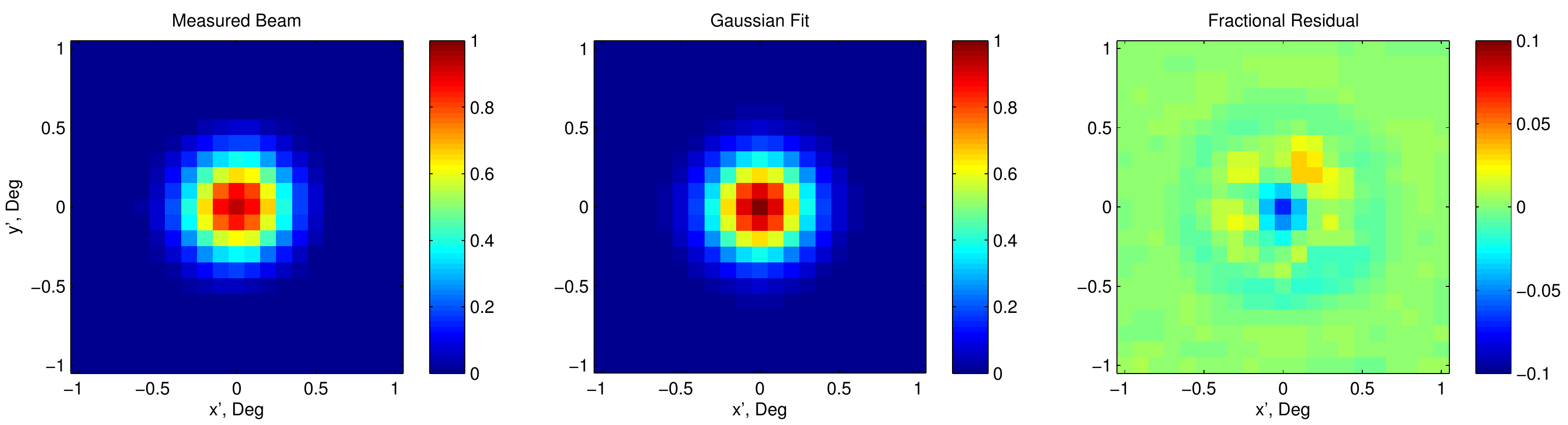}
   \end{tabular}
   \end{center}
   \caption{ \label{fig:farFieldBeam}
     Left: Example measured far-field beam pattern from \bicepp, linear scale. Middle: Gaussian fit 
     to measured beam pattern. Right: Fractional 
     residual after subtracting the Gaussian fit in the middle panel.  The residual represents 
     the portion of the beam that is not mitigated with the deprojection technique discussed in 
     Section~\ref{sec:simulation}.
     Note: The right-hand panel has a different 
     color scale than the left two panels.
   }
 \end{figure*}
\renewcommand{\tabcolsep}{4pt}
\begin{table*}[ht]
\begin{center}
\caption[Measured detector-pair beam parameters]{\label{table:beam}Measured detector-pair beam parameters for 
  \bicepp and \keck receivers for 2012 and 2013} 
\begin{tabular}[c]{ccccc}
\hline 
\multicolumn{2}{c}{\multirow{2}{*}{Receiver}}
& \multicolumn{3}{c}{Beam Parameter}\\
  \multicolumn{2}{c}{}& Beam Width, $\sigma_i$ (degrees) & Ellipticity Plus, $p_i$ ($+$) & Ellipticity Cross, $c_i$ ($\times$)\\
\hline \hline
\multicolumn{2}{c}{\bicep2}& $0.220\pm0.004\pm0.002$ & $0.01\pm0.03\pm0.02$ & $0.00\pm0.02\pm0.01$\\
\cline{1-5}
\multirow{6}{*}{Keck 2012} & Receiver 0&$0.216\pm0.003\pm0.002$ & $0.01\pm0.03\pm0.01$ & $0.01\pm0.02\pm0.01$ \\
& Receiver 1& $0.215\pm0.004\pm0.002$& $0.01\pm0.02\pm0.02$& $0.02\pm0.02\pm0.01$ \\
& Receiver 2& $0.217\pm0.004\pm0.002$& $0.01\pm0.03\pm0.01$& $0.01\pm0.02\pm0.02$ \\
& Receiver 3& $0.217\pm0.004\pm0.002$& $0.01\pm0.02\pm0.01$& $0.01\pm0.02\pm0.02$ \\
& Receiver 4& $0.217\pm0.004\pm0.002$& $0.01\pm0.02\pm0.01$& $0.02\pm0.02\pm0.02$ \\
& All receivers & $0.216\pm0.004\pm0.002$ & $0.01\pm0.02\pm0.01$ & $0.01\pm0.02\pm0.02$ \\
\cline{1-5}
\multirow{6}{*}{Keck 2013} & Receiver 0 & $0.218\pm0.004\pm0.002$& $0.01\pm0.03\pm0.01$& $0.01\pm0.02\pm0.02$ \\
& Receiver 1 & $0.215\pm0.004\pm0.002$& $0.01\pm0.02\pm0.02$& $0.01\pm0.02\pm0.02$ \\
& Receiver 2 & $0.218\pm0.004\pm0.002$& $0.01\pm0.03\pm0.02$& $0.01\pm0.02\pm0.02$ \\
& Receiver 3 & $0.218\pm0.005\pm0.003$& $0.01\pm0.03\pm0.02$& $0.00\pm0.02\pm0.02$ \\
& Receiver 4 & $0.211\pm0.002\pm0.002$& $0.00\pm0.02\pm0.02$& $0.01\pm0.02\pm0.02$ \\
& All receivers &$0.216\pm0.005\pm0.002$ & $0.01\pm0.03\pm0.02$ & $0.01\pm0.02\pm0.02$ \\
\hline
\end{tabular}
\end{center}
\end{table*} 

\subsubsection{Beam shape parameters}
\label{sec:beamshape}
To facilitate parameterization of the beams, we first define
a coordinate system that is fixed to the focal plane as projected onto the sky
for each of the \bicepp and \keck receivers.
We scan the telescope in azimuth and step in elevation, and convert the angular offset
between the boresight and the direction to the beam mapping source into this coordinate system.

A pixel, $P$, which has two orthogonally polarized detectors, 
is defined to be at a location $(r,\theta)$ from the boresight, $B$.
We define
$r$ as the radial distance away from the boresight and $\theta$ as the 
counterclockwise angle looking out from the telescope towards the sky from
the $\theta = 0^\circ$ ray (along the boresight), which is defined to be the great circle that
runs between Tile 1 and Tile 2 on the focal plane. 
Tiles are numbered counterclockwise looking directly down on the focal plane,
and Tile 1 and Tile 2 are physically located on the side of the focal plane that 
connects to the heatstraps to the sub-kelvin refrigerator. 

We then define an ($x'$,$y'$) coordinate
system locally for each pixel $P$. The positive $x'$ axis is defined to be
along the great circle that passes through the point $P$
that is an angle $-\theta$
from the $\hat{r}$ direction of the pixel (back toward $B$). 
The $y'$ axis is defined as the great circle
that is $+90^\circ$ away from the $x'$ axis. 
The ($x'$,$y'$) coordinate system is then 
projected onto a plane
at $P$\footnotemark[6]\footnotetext[6]{We note that while this coordinate system is used
consistently throughout this paper and its companions, 
a previous description of our detectors by~\cite{roger} used the notation ``horizontal''
and ``vertical'', which in this coordinate system corresponds to $y'$ and $x'$ respectively.}.

This coordinate system is fixed to the instrument and rotates on the sky with
the boresight rotation angle, also called the deck angle $K$,
and the angle at which each receiver is clocked with respect to the $K=0$ line, also
called the drum angle $K'$. 
For \bicepp, $K'=0$.  Each \keck receiver has a $K'$ rotated by $\sim72^\circ$ from 
its two neighboring receivers in the 2012 and 2013 configurations. 

A two-dimensional elliptical Gaussian is parametrized by six parameters:
the gain, the two-parameter position of the center of the beam, 
and three parameters that together describe the beam
width and the ellipticity. 
We fit a two-dimensional elliptical Gaussian profile 
to the main beam of each detector in \bicepp and the \keck in a flat sky approximation, 
according to
\begin{eqnarray}\label{eqn:gaussian}
B(\mathbf{x})=\frac{1}{\Omega}e^{-\frac{1}{2}(\mathbf{x}-\boldsymbol\mu)^T\Sigma^{-1}(\mathbf{x}-\boldsymbol\mu)}
\end{eqnarray}
where $\mathbf{x}$ is the location of the beam center, $\boldsymbol\mu$ is the origin, 
$\Omega$ is the normalization factor, and $\Sigma$ is the covariance matrix. 
A common parameterization for $\Sigma$ uses the widths along the major and 
minor axis, $\sigma_{maj}$ and $\sigma_{min}$, and
the rotation angle of the major axis 
away from the $x'$ axis, $\phi$.
Here we have
\begin{eqnarray}
\Sigma = R^{-1}CR,
\end{eqnarray}
where 
\begin{eqnarray}
C =
\left(\begin{array}{cc}
\sigma_{maj}^{2} & 0 \\
0 & \sigma_{min}^{2}
\end{array}\right),
\end{eqnarray}
and the rotation matrix, $R$, is described as
\begin{eqnarray}
R =
\left(\begin{array}{cc}
\cos \phi & \sin \phi \\
- \sin \phi & \cos \phi
\end{array}\right).
\end{eqnarray}

We then define the ellipticity as 
\begin{equation}
e = \frac{\sigma_{maj}^2-\sigma_{min}^2}
{\sigma_{maj}^2 + \sigma_{min}^2}.
\end{equation}
The ellipticity alone does not specify the direction of the major or minor axis.
As in the Systematics Paper, instead of using
$\sigma_{maj}$, $\sigma_{min}$, and $\phi$ to describe the beam,
we introduce the parameters $\sigma$, $p$, and $c$ to describe the 
beam width and ellipticity 
in the ``plus'' and ``cross'' directions respectively.
In this basis, we have
\begin{equation}
\sigma^2 = \frac{\sigma_{maj}^2+\sigma_{min}^2}{2} 
\end{equation}
and 
\begin{equation}
\Sigma =
\left(\begin{array}{cc}
\sigma^2 (1+p) & c \sigma^2 \\
c \sigma^2 & \sigma^2(1-p)
\end{array}\right).
\end{equation}

$p$ and $c$ are related to $\sigma_{maj}$, $\sigma_{min}$, and $\phi$ by
\begin{equation}
c = \frac{\sigma_{maj}^2 - \sigma_{min}^2}{\sigma_{maj}^2 + \sigma_{min}^2} \sin 2\phi ,
\end{equation}
and
\begin{equation}
p = \frac{\sigma_{maj}^2 - \sigma_{min}^2}{\sigma_{maj}^2 + \sigma_{min}^2} \cos 2\phi .
\end{equation}

An elliptical Gaussian with its major axis oriented along the $x$-axis
or the $y$-axis has a purely $+p$ or $-p$ ellipticity, while an elliptical Gaussian
oriented diagonally ($\pm45\deg$) has a purely $\pm c$ ellipticity.
The total ellipticity is
\begin{equation}
e = \sqrt{p^2+c^2}.
\end{equation}

Using this parameterization, differential beam parameters 
for a pair of co-located orthogonally polarized 
detectors may be defined simply by taking the difference 
between the beam parameters of the two detectors.
There is a straightforward relationship between parameters calculated in this way
and the first order terms of a power series expansion of a Gaussian.
This relates the modes that are removed from the real data through 
deprojection directly to the difference beam
parameters obtained from beam map measurements, as discussed in the Systematics Paper.

Figure~\ref{fig:farFieldBeam} shows an example map from \bicepp,
the elliptical Gaussian fit, and the fractional residual after subtracting the fit.
Beams and their fits for the \keck are similar~\citep{vieregg12}.    
Table~\ref{table:beam} shows 
the median value of the relevant beam parameters for each receiver in each experiment, 
as well as the detector-to-detector scatter (taken to be half of 
the width of the central 68\% of the distributon of median values for each detector)
and the median measurement uncertainty for individual detectors (the median over all detectors 
of half of the width of the central 68\% of the distribution of measurements for a given detector).

The measured values in the table come from beam maps made using a chopped thermal source in the far field. 
The per-detector pair parameters for \bicepp are calculated as a 
combination of the elliptical Gaussian fits to 24 beam maps
with equal boresight rotation coverage.
Each measurement of each 
detector's main beam must pass a set of criteria to be included in the final extraction 
of beam parameters, including
a check that the beam center was not near the edge of the mirror and excluding measurements
where the initial elliptical Gaussian fit failed.
The \keck beam parameters for 2012 and 2013 are obtained from independent
sets of beam maps taken in February 2012 and February 2013 respectively. 
The maps were taken using a chopped thermal source in the far field. 
We separately characterize each \keck receiver for 2012 and 2013 because of the changes made to
the receivers between the two observing seasons (see Table~\ref{table:receivers}).

The absolute detector gain calibration as well as each detector's absolute pointing information
come from calibrating against CMB~temperature maps from Planck~\citep{planckI,planckVI}, as discussed in
the Instrument Paper. 
We do not obtain measurements of the absolute gain of each detector from
the beam maps since we use the aluminum transition and a bright source. 
We also do not make the final pointing measurements of each detector using the beam map data; 
instead, we obtain them directly from CMB maps.

\begin{figure}[ht]
   \begin{center}
   \begin{tabular}{c}
   \includegraphics[height=8cm]{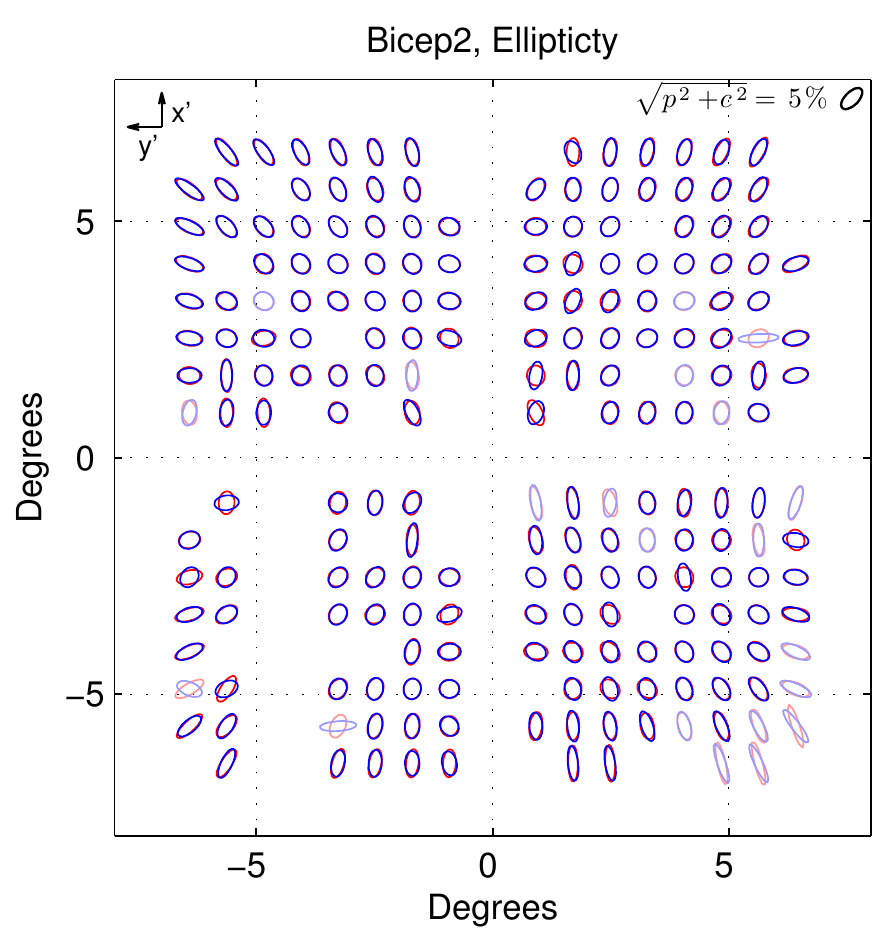}
   \end{tabular}
   \end{center}
   \caption{ \label{fig:ellip}  
     Per-detector beam ellipticity for \bicepp as projected onto the sky.
     The ellipticity for each detector has been exaggerated for visibility, as shown in the legend.
     $A$ and $B$ beams are
     shown in red and blue, respectively, and light colors show detectors that are not used in analysis.
   } 
   \end{figure}

\begin{figure*}[ht]
   \begin{center}
   \begin{tabular}{c}
   \includegraphics[height=14.3cm]{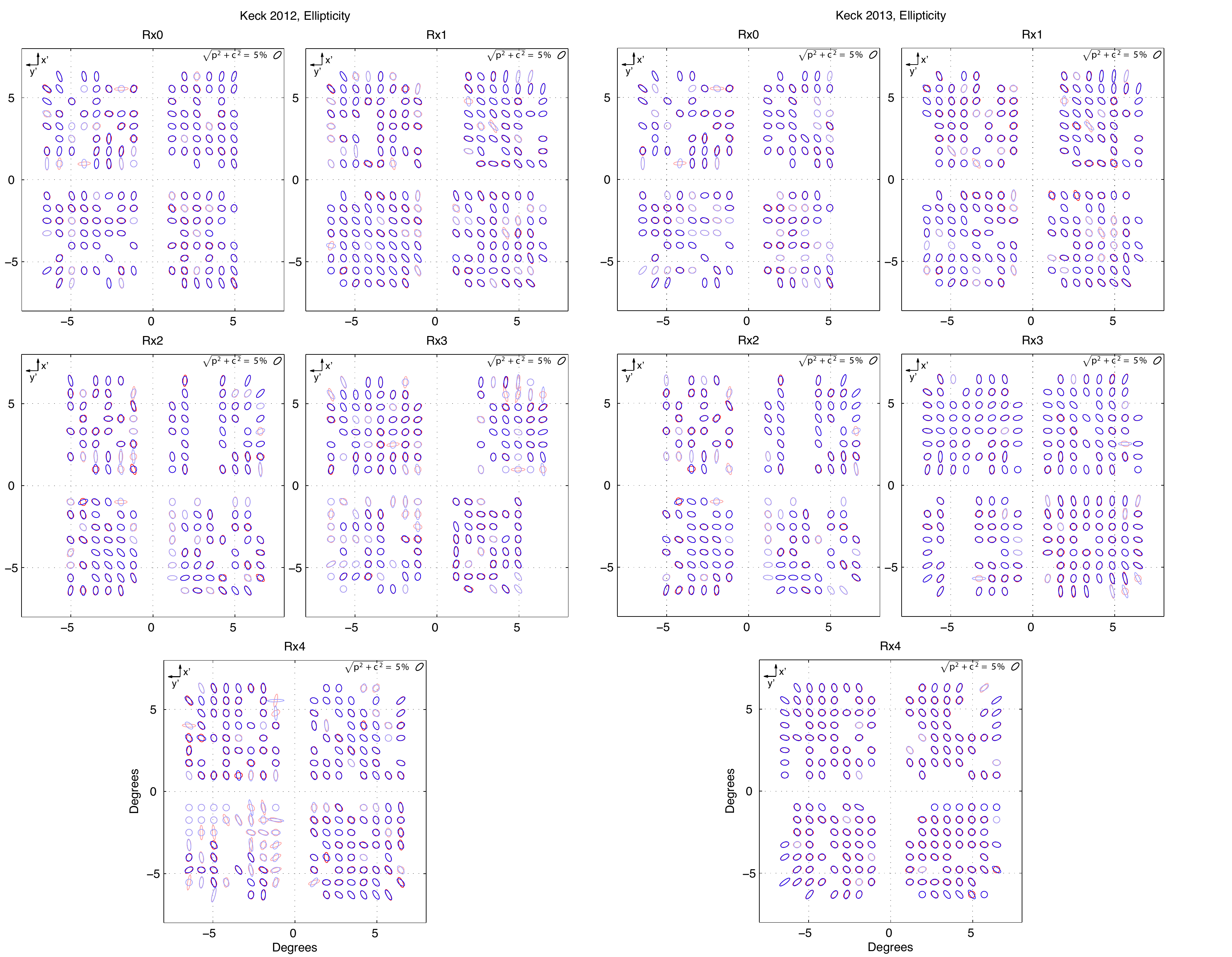}
   \end{tabular}
   \end{center}
   \caption{ \label{fig:ellipKeck}  
     Per-detector beam ellipticity for the \keck 2012 and 2013 configurations as projected onto the sky.
     The ellipticity for each detector has been exaggerated for visibility, as shown in the legend. 
     $A$ and $B$ beams are
     shown in red and blue, respectively, and light colors show detectors that are not used in analysis.
     Detectors in Receivers~0, 2, 
     and three of the four tiles on Receiver 1 are the same between 2012 and 2013, and the correlation
     between years for those receivers is evident.
   } 
   \end{figure*}

Figures~\ref{fig:ellip} and~\ref{fig:ellipKeck} 
show the distribution of each detector's
ellipticity across the focal plane for \bicepp and the \keck in its 2012 and 2013 configurations.
There are two effects that contribute to the observed ellipticity pattern across the 
focal plane.  First, because our optical design places its optimal focus on an annulus of 
detectors a median distance from the center of the focal plane, ellipticity is induced
for detectors near the edge of the focal plane.  This effect is predicted by our optical 
simulations.  Second, the beam steer effects that we observe in near field maps (described in 
Section~\ref{sec:nearfield}) also lead to enhanced ellipticity for detectors 
near the edges of tiles in the focal plane.  Although optical simulations do not predict 
the beam steer effects
seen in the near field, given the observed beam steer optical simulations do predict the
observed enhanced ellipticity in the far field.   The net effect 
in the far field is a combination of the two effects, which leads to detectors near the 
edges of tiles and near the edge of the focal plane displaying higher ellipticity.

\begin{figure*}[ht]
   \begin{center}
   \begin{tabular}{c}
   \includegraphics[height=5cm]{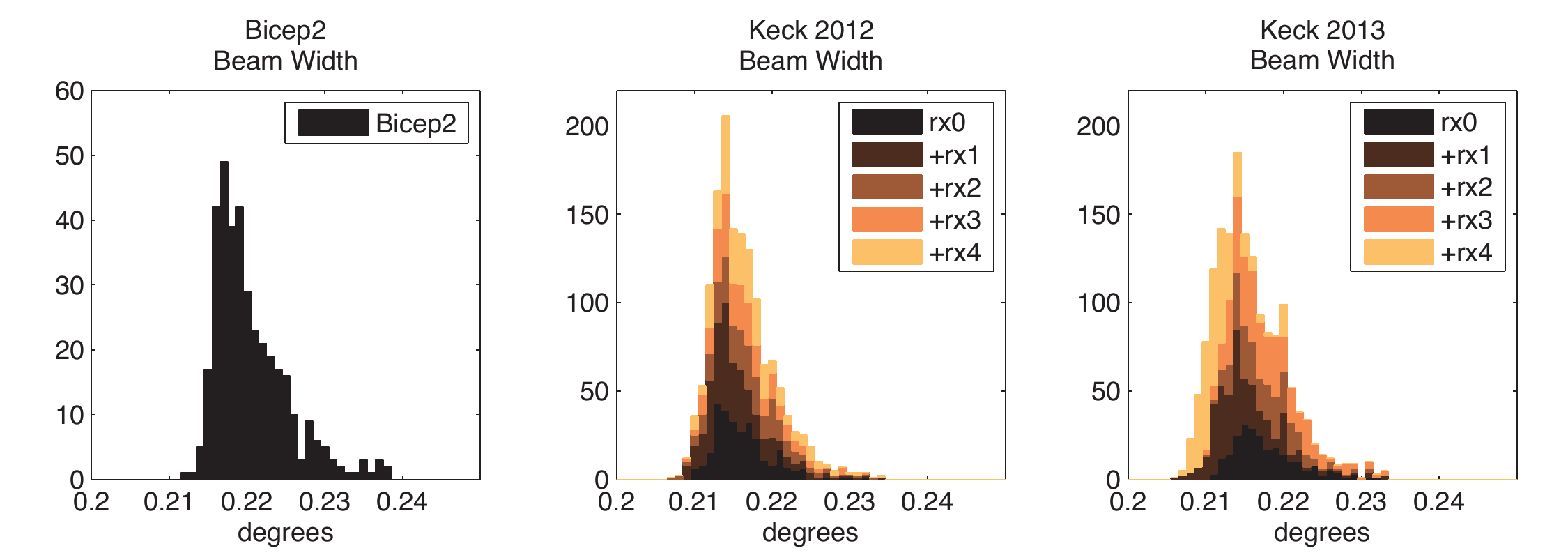}
   \end{tabular}
   \end{center}
   \caption{ \label{fig:sig_b2}  
     Per-detector beam width ($\sigma$) measurements for \bicepp (left-hand panel) and the \keck 2012 (middle panel) 
     and 2013 (right-hand panel).
   } 
   \end{figure*}

Figure~\ref{fig:sig_b2} shows the per-detector beam width $\sigma$ for \bicepp
and the \keck.  The median beam width over the focal plane is $0.220\pm0.002\pm0.001^\circ$ 
for \bicepp, $0.216\pm0.004\pm0.002^\circ$ for \keck receivers in 2012, 
and $0.216\pm0.005\pm0.002^\circ$ for \keck receivers in 2013, where the errors quoted are the 
detector-to-detector scatter followed by the measurement uncertainty for individual 
detectors (as in Table~\ref{table:beam}).
The beam widths for each \keck receiver are consistent
from one receiver to another.
The slight difference in the placement of the objective lens between \bicepp and 
the \keck could 
explain the difference in the beam widths between the \keck and \bicepp.


   \begin{figure}[ht]
   \begin{center}
   \begin{tabular}{c}
   \includegraphics[height=8cm]{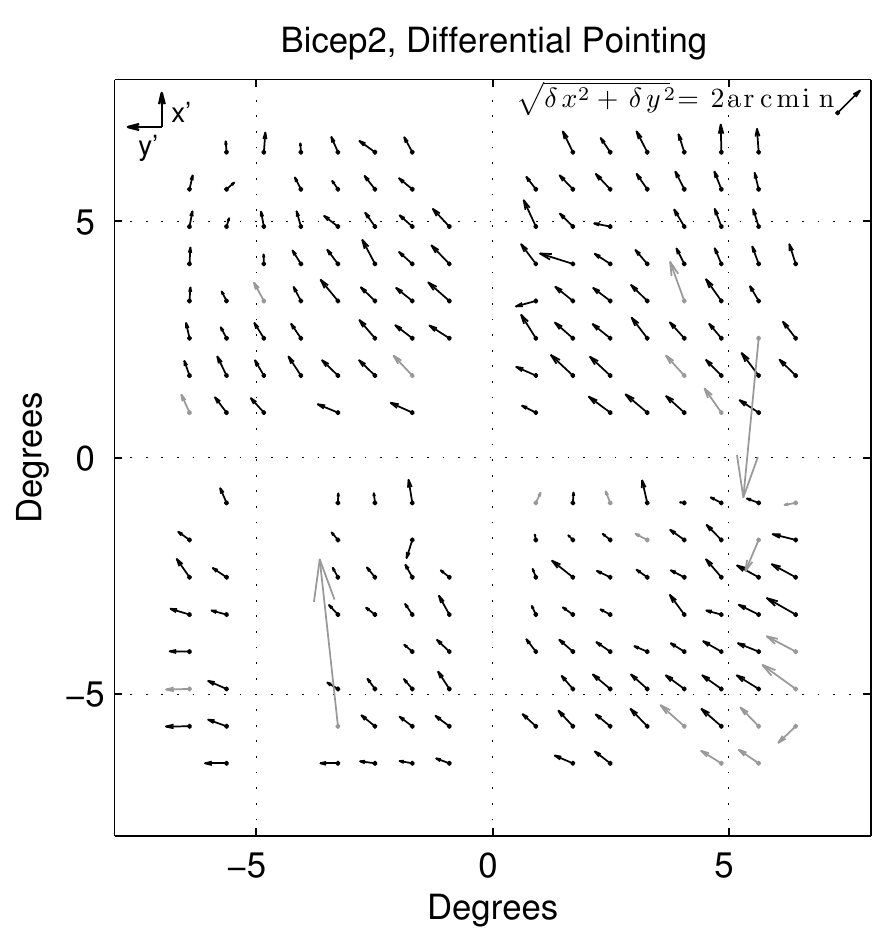}
   \end{tabular}
   \end{center}
   \caption{ \label{fig:diffPoint}  The differential pointing measured between orthogonally polarized, 
     co-located detector pairs, plotted in a focal plane layout for \bicepp.  
     Arrows point from the $A$ detector location to the $B$ detector location, and the length of the arrows 
     corresponds to the degree of mismatch multiplied by a factor of~20 for plotting.  Detector pairs with 
     grey arrows are not used in analysis.
   } 
   \end{figure}
   \begin{figure*}[ht]
   \begin{center}
   \begin{tabular}{c}
   \includegraphics[height=14.3cm]{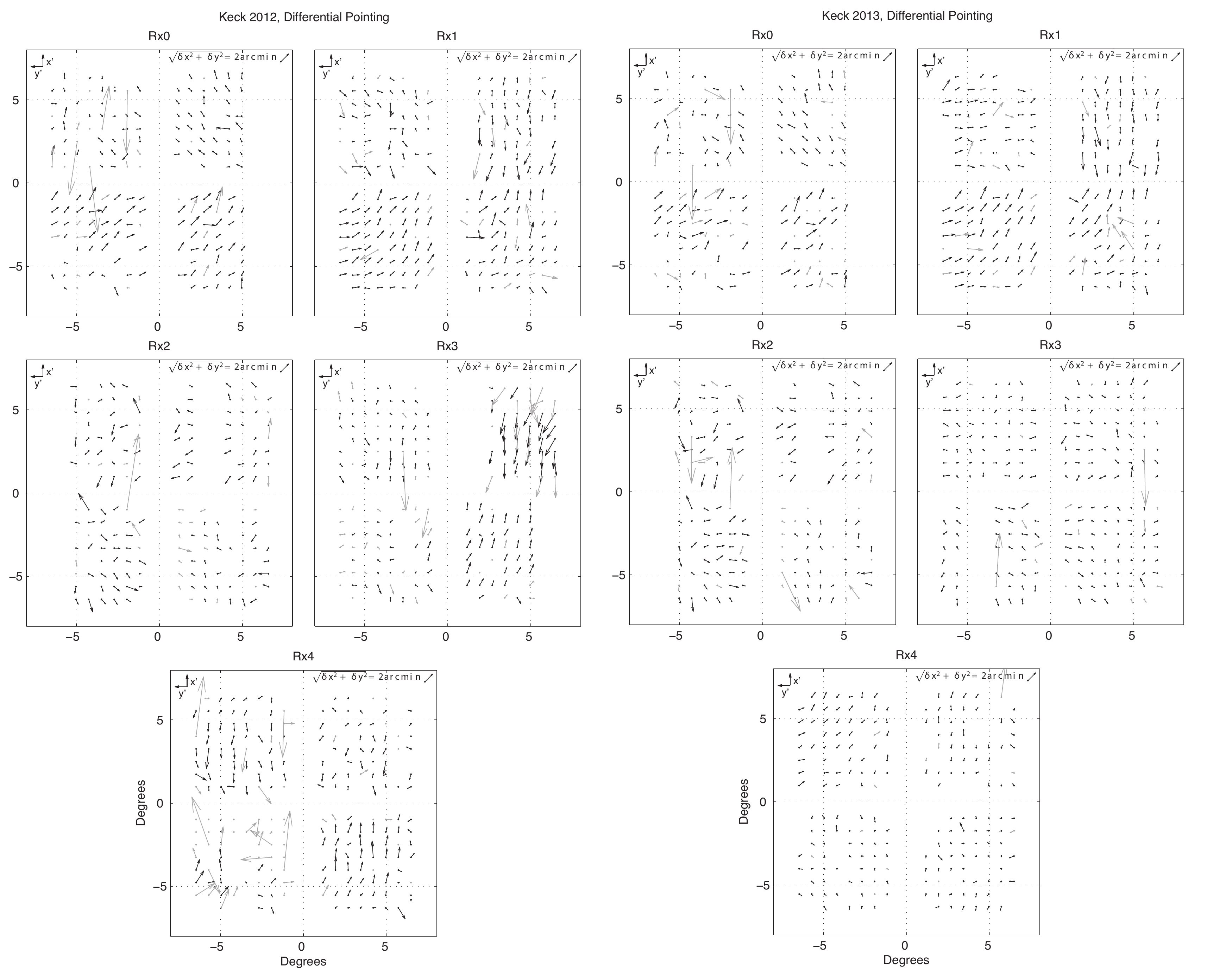}
   \end{tabular}
   \end{center}
   \caption{ \label{fig:diffPointKeck}  The differential pointing measured between orthogonally polarized, 
     co-located detector pairs, plotted in a focal plane layout for the \keck in its 2012 and 2013 configurations.  
     Arrows point from the $A$ detector location to the $B$ detector location, and the length of the arrows 
     corresponds to the degree of mismatch multiplied by a factor of~20 for plotting. Detector pairs with 
     grey arrows are not used in analysis. Detectors in Receivers~0, 2, 
     and three of the four tiles on Receiver 1 are the same between 2012 and 2013, and the correlation
     between years for those receivers is evident.  Receiver 4 received a new 
     focal plane in 2013 with reduced near-field differential pointing.
   } 
   \end{figure*}
   \begin{figure}[ht]
   \begin{center}
   \begin{tabular}{c}
   \includegraphics[height=8cm]{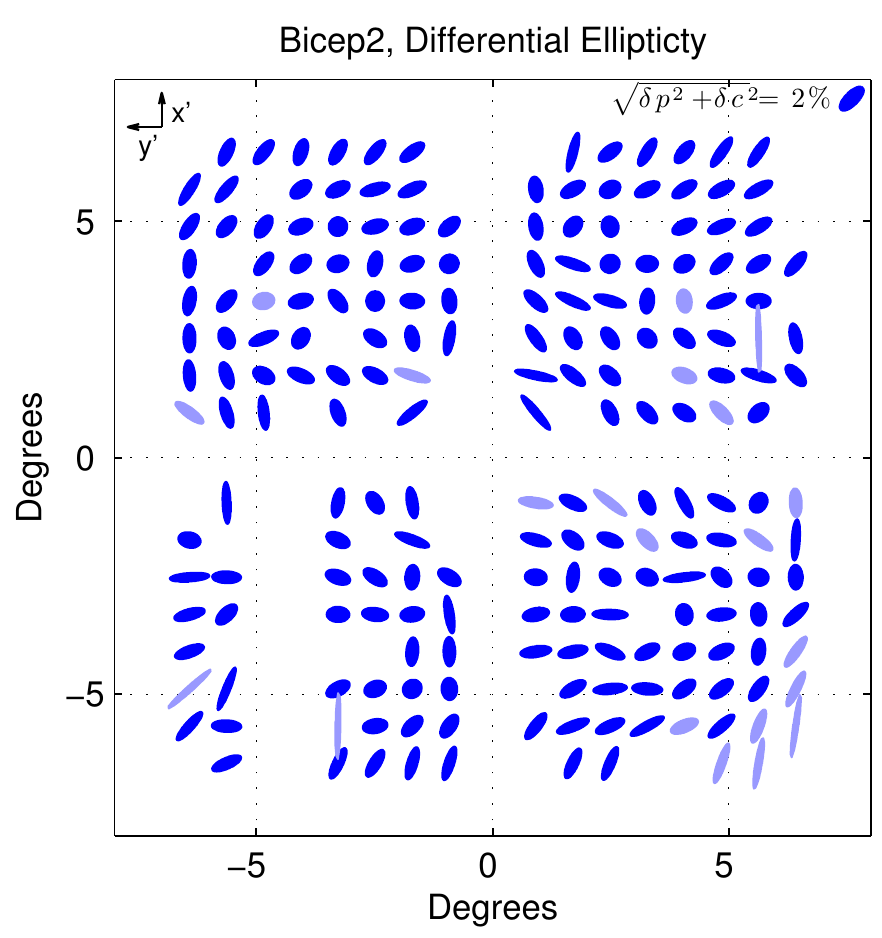}
   \end{tabular}
   \end{center}
   \caption{ \label{fig:diffEll} The differential ellipticity measured between orthogonally polarized, 
     co-located detector pairs, plotted in a focal plane layout for \bicepp.  The
     major axes of the ellipses are proportional to $\sqrt{\delta p^2 + \delta c^2}$, a measure of the magnitude of 
     the differential ellipticity.  A large fraction of
     detectors have beams whose ellipticity is well matched.  Light colored detector pairs are not used in analysis.  
     The differential ellipticity for each detector pair has been exaggerated for visibility, as shown in the legend.
   } 
   \end{figure}
   \begin{figure*}[ht]
   \begin{center}
   \begin{tabular}{c}
   \includegraphics[height=14.0cm]{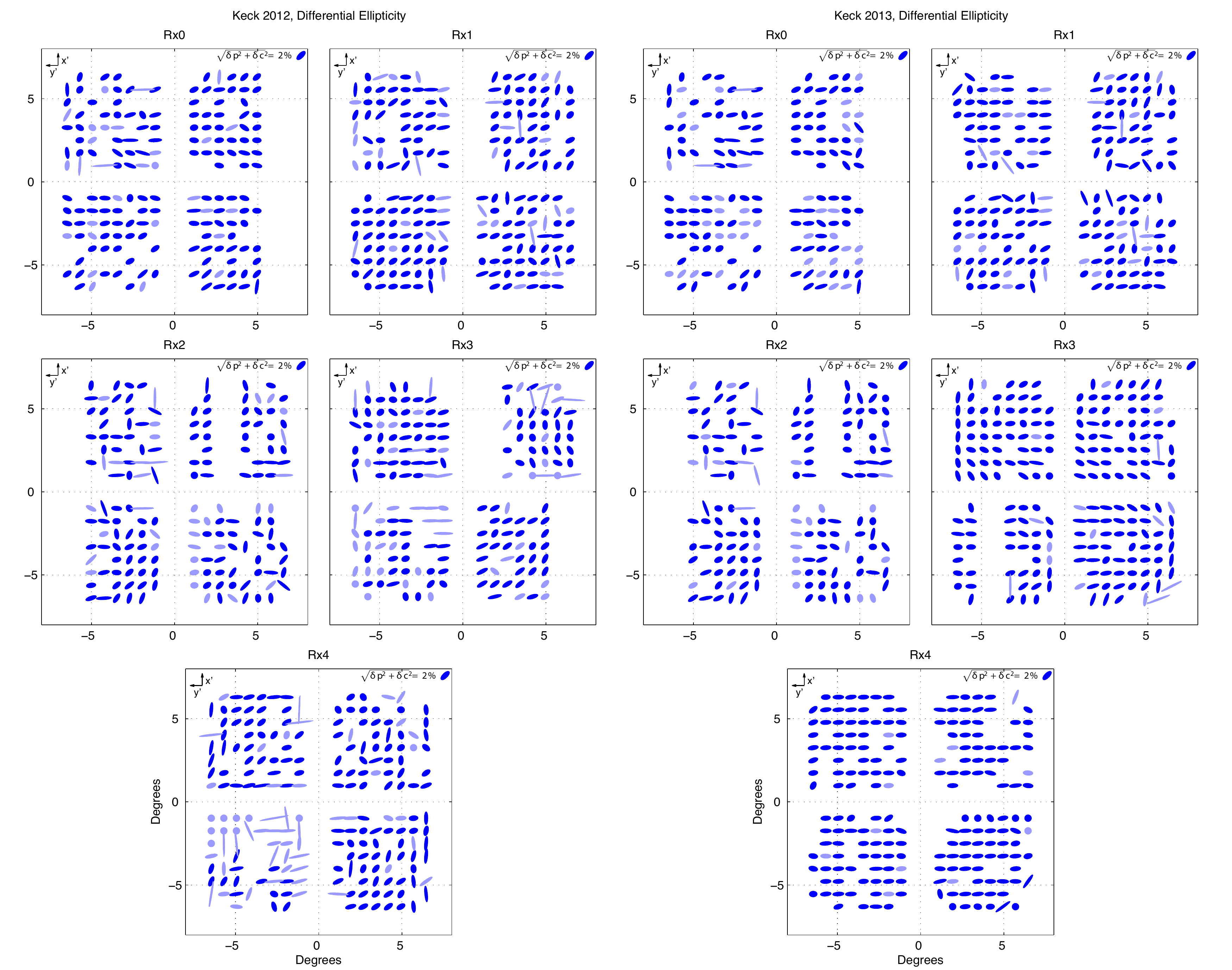}
   \end{tabular}
   \end{center}
   \caption{ \label{fig:diffEllKeck} The differential ellipticity measured between co-located detector pairs, plotted
     in a focal plane layout for the \keck 2012 and 2013. The
     major axes of the ellipses are proportional to $\sqrt{\delta p^2 + \delta c^2}$, a measure of the magnitude of 
     the differential ellipticity. 
     Light colored detector pairs are not used in analysis.    The differential ellipticity for each 
     detector pair has been exaggerated for visibility, as shown in the legend.
     Detectors in Receivers~0, 2, 
     and three of the four tiles on Receiver 1 are the same between 2012 and 2013, and the correlation
     between years for those receivers is evident.  
   } 
   \end{figure*}
   \begin{figure*}[ht]
   \begin{center}
   \begin{tabular}{c}
   \includegraphics[height=5cm]{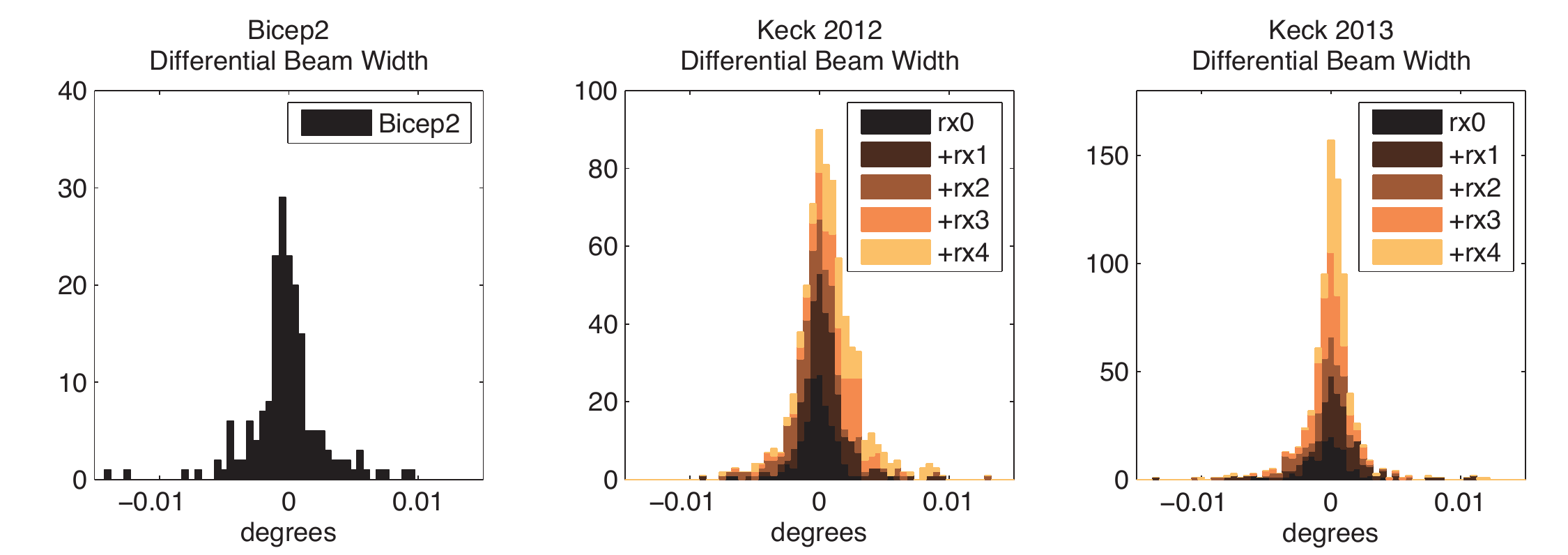}
   \end{tabular}
   \end{center}
   \caption{ \label{fig:diffWidth} The measured differential beam width between orthogonally polarized, 
     co-located detector pairs for \bicepp (left-hand panel) and the \keck 2012 (middle panel) and 2013 (right-hand panel).
   } 
   \end{figure*}

\subsubsection{Differential beam parameters} 
\label{sec:diffbeam}
We calculate differential beam parameters for a pair of co-located orthogonally polarized detectors by taking
the difference between the main beam parameters for each detector within the pair.
The differential beam parameters are calculated for each beam map measurement.
Receiver-averaged differential beam parameters 
for all detectors used in B-mode analysis are shown in Table~\ref{table:diffBeam}.  The uncertainties 
are calculated in the same way as in Table~\ref{table:beam}, described in Section~\ref{sec:beamshape}.
The scatter is dominated
by true pair-to-pair differences, not measurement repeatability.
The changes in the \keck beam parameters from 2012 to 2013 are explained by the focal plane
reconfigurations described in Table~\ref{table:receivers}.

Figures~\ref{fig:diffPoint} and~\ref{fig:diffPointKeck} 
show the measured differential pointing
and Figures~\ref{fig:diffEll} and~\ref{fig:diffEllKeck} show 
the measured differential ellipticity on a per-pair basis across the 
focal plane for \bicepp and the \keck.  
Figure~\ref{fig:diffWidth} shows
histograms of the measured differential beam width on a per-pair basis for \bicepp and the \keck. 

We observe a pattern of far-field differential pointing across the focal plane (see 
Figures~\ref{fig:diffPoint}~and~\ref{fig:diffPointKeck}) that is different for each receiver and
has no direct correlation with the observed 
near-field mismatch for a given focal plane, which also exhibits a pattern across the focal plane
(see Section~\ref{sec:nearfield}).
The source of the far-field differential pointing could be an interaction of the observed
near-field mismatch with possible imperfections in the optical system that would translate the near-field
mismatch into the far field (see Section~\ref{sec:nearfield}).
The observed differential pointing per pair for \bicepp was larger than that observed for \bicepo and much 
larger than optical
modeling of the telescope predicts (see Section~\ref{sec:design}). 

The differential beam width for all receivers is
small and does not have an observable pattern across the focal plane.
While there is a large spread of per-detector ellipticities across the focal plane, 
ellipticities within each detector pair are relatively well matched, so
the range of differential ellipticities for each detector pair is smaller than the
scatter in the ellipticities on a per-detector basis.
The differential ellipticities for \bicepp and \keck are 
larger at the edges of each tile, evident in
Figures~\ref{fig:diffEll} and~\ref{fig:diffEllKeck}.


\begin{table*}[ht]
\begin{center}
\caption[Measured differential beam parameters]{\label{table:diffBeam}Measured pair-difference beam parameters for 
  \bicepp and \keck receivers for 2012 and 2013} 
  \begin{tabular}[c]{llccccc}
    \hline 
    \multicolumn{2}{c}{\multirow{3}{*}{Receiver}}
    & \multicolumn{5}{c}{Differential Beam Parameters}\\
    \multicolumn{2}{c}{}& 
    Differential X Pointing &
    Differential Y Pointing &
    Differential Beam Width &
    Diff. Plus Ellipticity & 
    Diff. Cross Ellipticity\\
    \multicolumn{2}{c}{}& 
    $dx_i$ (arcminutes) &
    $dy_i$ (arcminutes) &
    $d\sigma_i$ (degrees) &
    $dp_i$ ($+$) & 
    $dc_i$ ($\times$)\\
    \hline \hline
    \multicolumn{2}{c}{\bicep2}
    & $0.81\pm0.29\pm0.14$ & $0.78\pm0.35\pm0.14$ 
    & $0.000\pm0.001\pm0.001$ 
    & $-0.002\pm0.013\pm0.011$ & $-0.003\pm0.012\pm0.005$ \\
    \cline{1-7}
    \multirow{5}{*}{Keck 2012} & Receiver 0 
    &$0.60\pm0.34\pm0.07$ & $-0.27\pm0.63\pm0.11$ 
    &$0.000\pm0.001\pm0.001$
    & $-0.011\pm0.009\pm0.003$ & $-0.004\pm0.008\pm0.003$\\
    & Receiver 1 
    & $-0.025\pm0.54\pm0.09$ & $-0.48\pm0.45\pm0.04$ 
    & $0.001\pm0.002\pm0.001$ 
    & $-0.007\pm0.011\pm0.002$ & $-0.009\pm0.006\pm0.002$\\
    & Receiver 2 
    & $-0.25\pm0.54\pm0.09$ & $0.14\pm0.69\pm0.09$ 
    & $-0.001\pm0.002\pm0.001$ 
    & $-0.007\pm0.015\pm0.002$ & $-0.003\pm0.009\pm0.003$\\
    & Receiver 3 
    & $-0.39\pm1.68\pm0.06$ & $-0.12\pm0.35\pm0.05$ 
    & $0.001\pm0.002\pm0.001$ 
    & $-0.007\pm0.015\pm0.002$ & $-0.007\pm0.008\pm0.002$\\
    & Receiver 4 
    & $-0.04\pm1.21\pm0.08$ & $-0.06\pm0.33\pm0.04$ 
    & $0.002\pm0.002\pm0.001$ 
    & $-0.002\pm0.016\pm0.003$ & $-0.009\pm0.009\pm0.002$\\
    \cline{1-7}
    \multirow{6}{*}{Keck 2013} 
    & Receiver 0 
    & $0.59\pm0.39\pm0.07$ & $-0.26\pm0.70\pm0.06$ 
    & $0.000\pm0.001\pm0.001$ 
    & $-0.011\pm0.009\pm0.002$ & $-0.004\pm0.009\pm0.002$\\
    & Receiver 1 
    & $0.03\pm1.12\pm0.06$  & $-0.57\pm0.56\pm0.05$  
    & $0.001\pm0.002\pm0.001$ 
    & $-0.010\pm0.016\pm0.002$ & $-0.009\pm0.006\pm0.002$\\
    & Receiver 2 
    & $-0.26\pm0.52\pm0.12$ & $0.10\pm0.69\pm0.11$ 
    & $-0.001\pm0.002\pm0.001$ 
    & $-0.007\pm0.016\pm0.003$ & $-0.003\pm0.009\pm0.003$\\
    & Receiver 3 
    & $-0.09\pm0.32\pm0.12$ & $-0.09\pm0.47\pm0.11$ 
    & $0.000\pm0.001\pm0.001$ 
    & $-0.001\pm0.013\pm0.004$ & $-0.002\pm0.009\pm0.003$ \\
    & Receiver 4 
    & $0.21\pm0.46\pm0.04$ & $-0.14\pm0.35\pm0.02$ 
    & $0.000\pm0.001\pm0.001$ 
    & $-0.020\pm0.005\pm0.002$ & $-0.002\pm0.003\pm0.001$ \\
    \hline
  \end{tabular}
\end{center}
\end{table*}


\subsubsection{Comparison with optical models} 
\label{sec:simCompare}
\begin{figure}[ht]
   \begin{center}
   \begin{tabular}{c}
   \includegraphics[height=6.6cm]{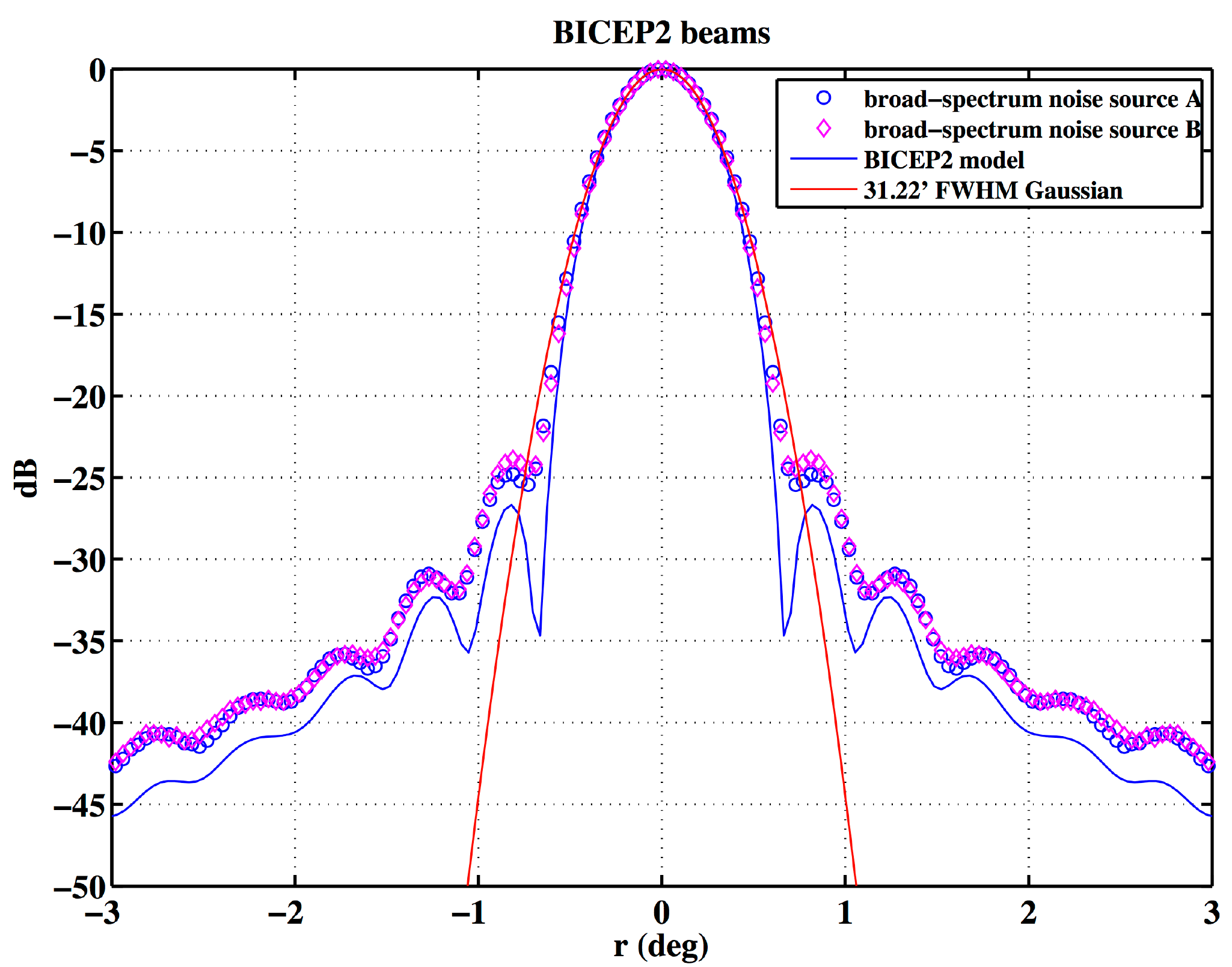}
   \end{tabular}
   \end{center}
   \caption{ \label{fig:zemaxCompare} Cross section of \bicepp beams (averaged over all pairs of detectors
     used in analysis) compared with Zemax physical optics 
     simulations and
     a Gaussian fit.  The agreement between the measured beams and the simulation is good, and the data are well-fit by
     a Gaussian model near the peak.
   } 
   \end{figure}

Figure~\ref{fig:zemaxCompare} shows a comparison between beams averaged over all pairs of detectors used in
analysis \bicepp
and the results of the Zemax physical optics model discussed in Section~\ref{sec:sim}.  Also shown is
a Gaussian fit to the beams.  The cross section
shown is taken along the scan direction, which is aligned with the horizon. 
The Zemax simulation is monochromatic, which gives rise to the sharp nulls in the Airy rings that are smoothed out
in the real data due to the wider bandpass of the detectors. 
Otherwise, the agreement between the measured beams and simulation is good. 

\subsection{Far sidelobes}
\label{sec:farsidelobes} 
Far sidelobes of the telescope can potentially see the bright Galactic plane, features on the ground, or 
emission from the Moon.  
The telescope ground shield systems for \bicepp and the \keck
were designed to mitigate contamination from the ground.

The ground shielding system, described in~\cite{takahashi10} and in Section~\ref{sec:shield}, 
has two main components.  First, there is a co-moving absorptive forebaffle 
that rotates around the boresight with the telescope and intersects beams at $\sim15^\circ$ from the peak.  
Second, there is a fixed reflecting ground shield.  The additional loading
on the detectors due to the forebaffle was measured to be $\sim$3--6~$\mathrm{K_{CMB}}$ in \bicepp 
and $\sim$5--10~$\mathrm{K_{CMB}}$ in the \keck.  The lowest loading was found for edge 
pixels and the highest loading was found for central pixels.  This is higher than the measured \bicepo value 
of $\sim$2~$\mathrm{K_{CMB}}$~\citep{takahashi10}.  
The origin of this coupling
is attributed to a combination of scattering off the foam window, shallow-incidence reflections off the 
inner wall of the cold telescope tube, and residual out-of-band coupling.  For the 2014 observing season,
we implemented absorptive corrugations inside the telescope tube of each \keck receiver to reduce 
shallow-incidence reflections off the telescope tube.  After the corrugations were installed, the 
total loading due to the forebaffle was reduced to 3~$\mathrm{K_{CMB}}$ for the 2014 observing season.

We measure the far-sidelobe response using a broad-spectrum noise source, described 
in Section~\ref{sec:rps}, with fixed polarization.  The source is mounted on a mast and sits
$\sim10$~m from the telescope so that the far sidelobes can be mapped with no flat mirror installed
and with a very bright source.
We achieve $\sim70$~dB of dynamic range by performing the measurements at two 
different source brightnesses to achieve
the signal-to-noise required to measure dim features far from the main beam while retaining 
the ability to measure the 
main beam itself without significant gain compression or instability.  

In \bicepp, we observe that while there are no sharp features in the far-sidelobe regime (defined as having a 
geometry such that it could see
Galactic emission during regular CMB observations), there is some power that is spread diffusely through
the far-sidelobe region.  Most of this power is intercepted by the forebaffle at $> 15^{\circ}$ from the main beam. 
We integrate the total power in a typical beam profile to quantify the fraction of the power found outside of 
a given angle from the main
beam center.  We find that for a typical detector, 
less than 0.1\% of the total integrated power is found outside of $25^\circ$ from the main beam for \bicepp
with the co-moving forebaffle installed.  We have mapped the far sidelobes of
the \keck and plan to perform a similar analysis.

To verify that the total power in the far sidelobes that intersects the forebaffle 
matches the amount of additional loading on the detectors due to the forebaffle, we 
make maps of the far-sidelobe response both with and without the forebaffle installed.  We then
measure the amount of far-sidelobe power that was intercepted by the forebaffle and compare it 
to the measured forebaffle loading on the detectors.  For \bicepp, 
the fractional amount of power intercepted by the forebaffle averaged 
across the focal plane is 0.7\%, corresponding to 3~$\mathrm{K_{CMB}}$ --- consistent with the measured excess loading 
due to the forebaffle.

\subsection{Polarization angle and cross-polar beam response}
\label{sec:polangle}

A key advantage of the \bicepp/\keck experimental approach is the ability to characterize the polarization
angles, $\psi$, and cross-polar response, $\epsilon$, 
of each detector to high precision using ground-based calibrators. 
To calculate the cross-polar response, we first find the polarization angle that maximizes 
the total integrated amplitude of each detector's response and then compare that amplitude 
with the amplitude when the radiation's polarization is rotated by $90^\circ$ from that angle.
This can be thought of as the monopole or gain term in the cross-polar response.

Precise characterization is
made possible by
\bicepp/\keck's relatively short far-field range (beyond 70~m). 
Polarization angle calibration is
important for constraining potential systematics. Large systematic uncertainty in the polarization orientation
of the detectors with respect to the sky would lead directly to E-to-B leakage, resulting in false EB
correlation. Because E and T are correlated, this would also result in false TB correlation. 

The procedure we use to make polarization maps in our B-mode analysis 
only requires precise measurement of
per-detector polarization angles, not of the overall rotation.
We use correlations seen in 
the CMB itself via a self-calibration procedure using TB and EB correlations~\citep{keating} to 
estimate the overall rotation angle.  This CMB calibration procedure indicates a coherent rotation 
of $\sim 1^{\circ}$ for \bicepp, which is then removed in the B-mode analysis as described in the Systematics Paper.

We derive a benchmark for polarization angle
measurement precision driven by systematics contamination requirements for a measurement of the BB spectrum 
that does not use the CMB self-calibration technique. Using the same technique presented 
in~\cite{takahashi10} for \bicepo, we find a benchmark of $\Delta\psi < 0.7^\circ$, corresponding to $r<0.01$.  
A more stringent benchmark of $\Delta\psi < 0.2^\circ$ 
is required by the desire to measure the EB spectrum with high precision to test proposed cosmological 
mechanisms that generate a non-zero EB spectrum, such as cosmic birefringence~\citep{carroll}.
The cross-polar response enters in analysis as a small adjustment to the overall polarization gain,
but cannot create any false B-mode signal.

\bicepp and the \keck use two different methods for determining polarization angles, which 
allows us to check for consistency between measurements and to look for systematics in the measurements 
themselves.
The first method uses a thin rotating dielectric
sheet placed directly above the vacuum window. 
The second method involves observing a rotating polarized 
source in the far field.  In the following sections, we will discuss results from each method.

\subsubsection{Dielectric sheet calibration}
\label{sec:dsc}
   \begin{figure}[ht]
   \begin{center}
   \begin{tabular}{c}
   \includegraphics[height=6cm]{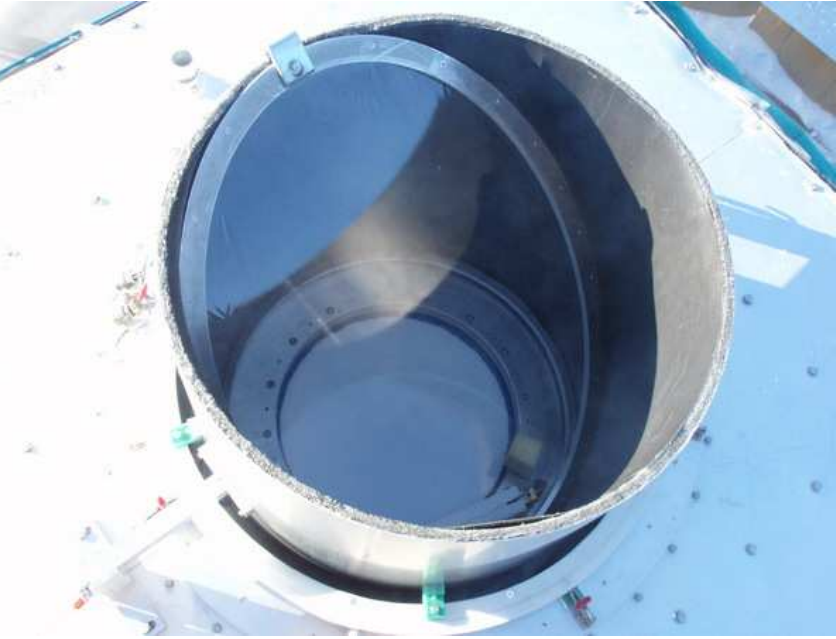}
   \end{tabular}
   \end{center}
   \caption{ \label{fig:dsc} A picture of the dielectric sheet calibrator installed on the \bicepo telescope.  We 
     used this calibrator to measure the polarization angle and cross-polar response of \bicepp as well.
   } 
   \end{figure}
   \begin{figure}[ht]
   \begin{center}
   \begin{tabular}{c}
   \includegraphics[height=4.3cm]{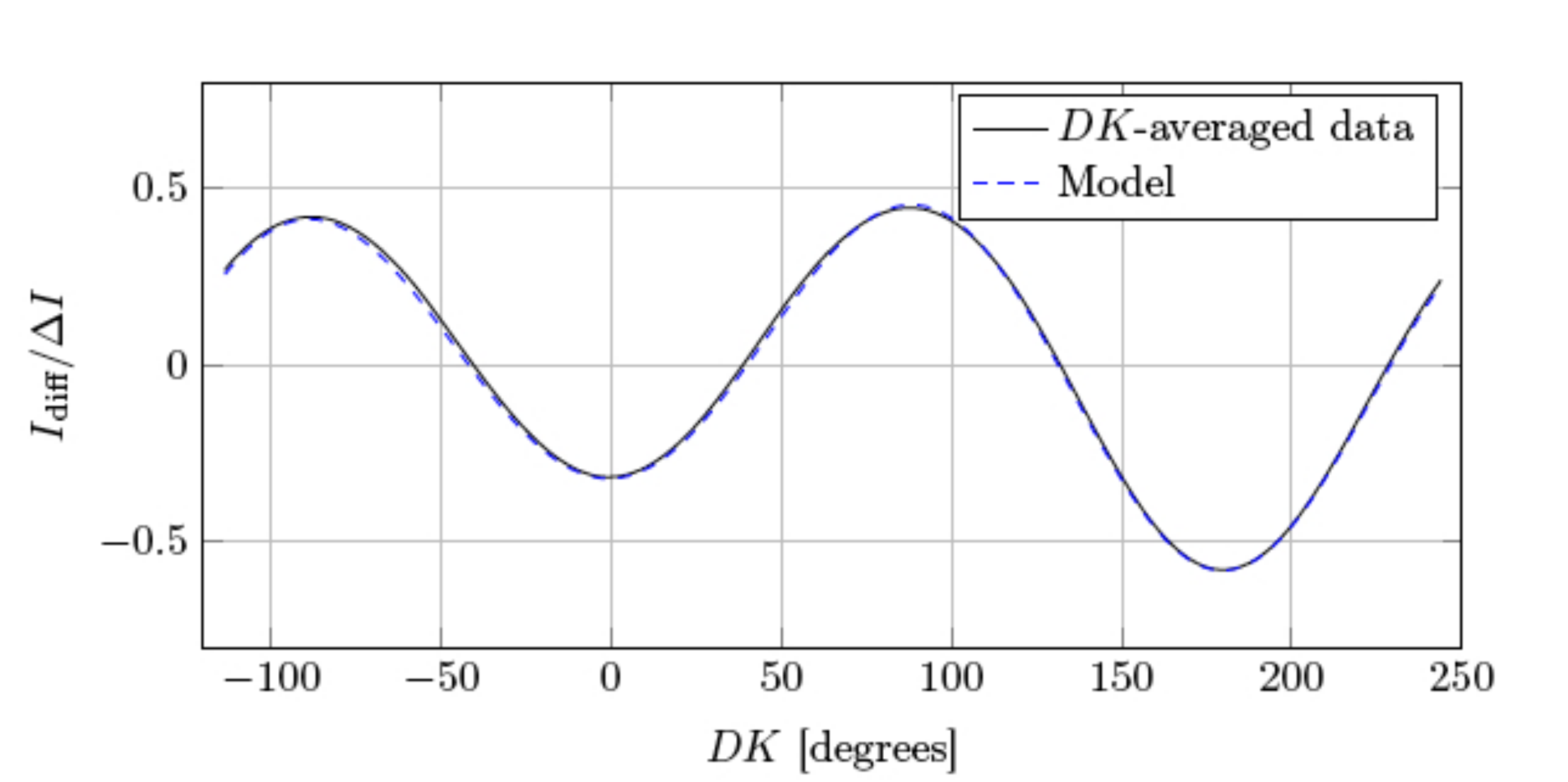}
   \end{tabular}
   \end{center}
   \caption{ \label{fig:yukiCal} The dielectric sheet calibrator pair-difference amplitude for a 
     typical pair of detectors
     in \bicepp ($I_{\mathrm{diff}}$), 
     normalized by the difference in temperature between the warm absorber and the sky ($\Delta I$), 
     plotted as a function of boresight rotation angle (DK).  
     The fitted model is also plotted.
   } 
   \end{figure}
The dielectric sheet calibrator (DSC) is a thin plastic film oriented at a $45^\circ$ angle to the optical axis
of the telescope, as shown in Figure~\ref{fig:dsc}.  The telescope is free to rotate about its boresight with 
respect to the thin film.  The film acts as a partially polarized beam splitter, preferentially 
reflecting one polarization of
the beam into the warm absorptive lining around the splitter and transmitting the other polarization
preferentially to the cold sky.  The DSC is essentially a beam-filling polarized source with a brightness
that scales with the difference in temperature between the sky and ambient.  
By rotating the telescope about its boresight and keeping the DSC fixed, 
we measure the polarization response of each detector as a function of boresight angle.
This technique is fast and precise for relative angles, but is sensitive to the exact alignment
of the calibrator with respect to the focal plane.  An identical technique was used for 
\bicepo~\citep{takahashi10}.

To make measurements of the polarization angle, 
the DSC is installed in place of the forebaffle directly above the vacuum window of the
telescope. Because a substantial fraction of the beam is transmitted to the sky, data are acquired 
only when the weather is good to avoid atmospheric noise in the measurement. 
A film thickness and index are chosen to provide the requisite signal-to-noise
while avoiding gain instability in the detectors. 
The detectors are then biased onto either the titanium or aluminum superconducting transition.
Before acquiring the calibration scan, the telescope is dipped in elevation to provide an unpolarized
signal modulation from which a relative gain correction between $A$ and $B$ detectors is derived. Scans are
acquired by counter-rotating in boresight rotation and azimuth with the telescope pointed at zenith. 
The counter-rotation fixes the beam location on the
sky while the calibrator (attached to the azimuth axis but not the boresight rotation axis) 
rotates about the boresight of the telescope.
   \begin{figure*}[ht]
   \begin{center}
   \begin{tabular}{c}
   \includegraphics[height=6cm]{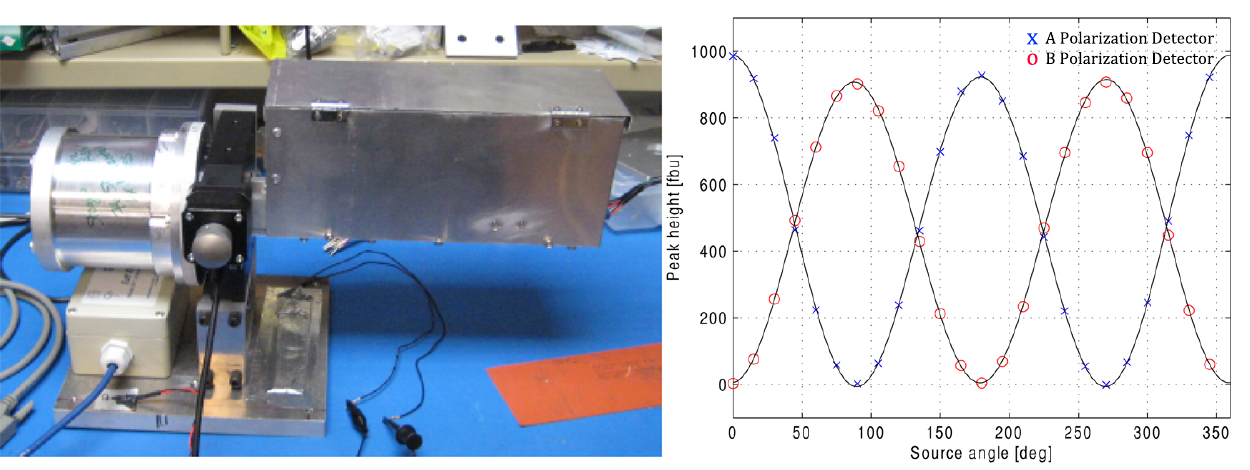}
   \end{tabular}
   \end{center}
   \caption{ \label{fig:rpsResponse} Left: The rotating polarized amplified thermal broad-spectrum noise source 
     used for polarization characterization. Right: Polarization modulation vs. source angle of an example 
     detector pair from \bicepp, measured
     using the rotating polarized source.
   } 
   \end{figure*}

An example of the periodic modulation 
of the detector pair-difference signal observed under rotation of the DSC
for \bicepp is shown in Figure~\ref{fig:yukiCal}.
The geometric model that we use to extract the polarization angle for a given detector from 
the observed modulation~\citep{takahashi10} relies 
on a few externally measured quantities: the tilt of the dielectric sheet, the
sheet material properties including the index of refraction and thickness, and the incident angle of
each of the detectors on the film. The tilt of the sheet was measured with respect to gravity with a digital
level before and after each scan and was close to $45^\circ$.  The lateral tilt
across the surface of the dielectric sheet was also measured with
a digital level.  Between installations of the calibrator, the value was observed to change
by as much as $1.5^\circ$, but during a measurement it was repeatably measured to <$0.02^\circ$.
The polarization angle is a weak function of both the index of refraction and the sheet 
thickness.  The incident angles of the detectors are taken from detector
centroid fits in the far field.  

With these external inputs accurately
measured, the model leaves only two free parameters. The first is the polarization angle $\psi$ of each 
detector. The second free parameter is the amplitude of the
signal ($\Delta I$), which is proportional to the difference in temperature between the absorptive lining and the
sky temperature at zenith and is a nuisance parameter.  The amplitude, normalized
by the brightness difference between the warm absorber and the sky, is extremely well-matched 
to the model of the polarized signal expected from the dielectric sheet.

For \bicepp, polarization angles are derived from a total of five independent measurements 
acquired between August 2010 and December 2012.  
The first three measurements used a 2~mil thick Mylar film and the final two measurements
were taken with a thinner 1~mil sheet. To combine the results, weights are derived
from the inverse variance of the residual after subtracting the fitted model. 
Using these measurements, the per-detector polarization angles have a statistical error of <$0.2^\circ$.
In the \bicepp B-mode analysis, we adopt the per-detector polarization angles from the
DSC for use in making polarization maps.

We took first measurements with the DSC for \keck receivers after the 2013 observing season, and plan 
to follow up with more complete measurements in future seasons.

\subsubsection{Rotating polarized source measurements}
\label{sec:rps}
We also measure the polarization angle and the cross-polar response of each detector 
using a rotating
polarized broad-spectrum noise source that we have developed
for use with \bicepp and the \keck, shown in Figure~\ref{fig:rpsResponse}, and called the Rotating Polarized 
Source (RPS)~\citep{bradford}.
The initial version of the broad-spectrum noise source emitted radiation in the 140--160~GHz range, 
designed to cover the passband of \bicepp and the initial installation of \keck receivers.  
Subsequently, the source was retrofitted to cover the frequency bands of \keck receivers at 95~GHz and 
220~GHz.
A $50~\Omega$ load provides room-temperature
thermal noise at the input of the first stage of amplification (80~dB).  A series of frequency multipliers, 
amplifiers, and filters bring the output frequency to the desired range (140--160~GHz for use with 150~GHz receivers). 
Linearly polarized 
radiation is emitted by a 15~dBi horn antenna, and is further polarized by a free-standing wire grid, 
yielding cross-polar leakage of the source $< 0.03$\%.
Two variable attenuators in series 
allow for control of output power over a large dynamic range, making the source useful for multiple applications,
including polarization measurements, far-field beam mapping, and
sidelobe mapping with the source closer to the receiver.  A microwave switch chops the source 
at $\sim18$~Hz. 
For RPS measurements, 
the entire source is mounted on a stepped rotating stage and 
has a total positional repeatability~$< 0.01^{\circ}$.

To map the response of every detector as a function of polarization angle
incident on the detector, we set the polarized source to a given polarization angle
and scan in azimuth over the source over a tight elevation range to obtain beam maps of
one physical row of detectors on the focal plane at one polarization angle.  
We then repeat this measurement in steps of 
$15^\circ$ in source polarization angle over a full $360^\circ$ range.  
After completing all source polarizations for a given row of detectors,
we move to the next row of detectors and repeat the sequence.  We repeat the entire set of measurements 
at two distinct boresight rotation angles as a consistency check.
An example of the polarization modulation as a function of source angle for one pair of \bicepp
detectors is shown in Figure~\ref{fig:rpsResponse}.  

We perform a five-parameter least squares fit to the detector 
response as a function of angle to extract a polarization angle and 
cross-polar response for each detector.  The model is described as:

\begin{equation} A\left(\cos{(2(\theta+\psi))}+\frac{1+\epsilon}{1-\epsilon}\right) \left(C\cos{(\theta+\phi)}+1\right), \label{eq:rps_response} \end{equation}
where $\theta$ is the angle of the source, 
$\epsilon$ is the cross-polar response of each detector, $\psi$ is the polarization angle of each detector, 
$A$ is the amplitude of the source, 
and $C$ and $\phi$ are the amplitude and phase of a source collimation term, common across all detectors
and describing any misalignment between the source rotation axis and the source alignment axis.  The source 
collimation misalignment gives rise to a sinusoidal response with a period of $360^\circ$, 
while the polarization modulation has a period of $180^\circ$.  This allows us to separate the two effects,
since the parameters are not degenerate.

The cross-polar response is very low for detectors in \bicepp, 
typically $\sim0.4$\%, with less than 10\% of detectors showing cross-polar response greater than 1\%.
This is consistent with the known level of crosstalk between two detectors in a given 
polarization pair and the level of direct island coupling in the detectors, discussed in the 
Instrument Paper and the Detector Paper.

\subsubsection{Polarization beam characterization}
\label{sec:polbeam}
In Section~\ref{sec:rps} we described our measurements of the 
monopole term in the cross-polar response and the polarization angle of each detector pair using the RPS. 
We can also use 
RPS measurements to investigate higher-order terms of the cross-polar response, which 
lead to E-to-B leakage. 
The higher-order response can be expressed by defining $T$, $Q$, and $U$ beams for each detector, 
which we call $B_T$, $B_Q$, and $B_U$. 
For a sky signal with linear polarization expressed in the detector $Q$/$U$ coordinate system, the 
full detector response is given by an integral of the $T$, $Q$, and $U$ beams over solid angle,
\begin{equation}
  \int{\left[ B_T(\mathbf{x}) T(\mathbf{x}) + B_Q(\mathbf{x}) Q(\mathbf{x}) + B_U(\mathbf{x}) U(\mathbf{x}) \right] d\Omega}.
\end{equation}

To define the detector $Q$ and $U$ axes, we adopt the convention that the integral
of $B_U$ is zero, \textit{i.e.} $B_U$ 
has no monopole component. This choice decouples the description of $B_Q$ and $B_U$
from absolute calibration of the detector polarization angles. An 
alternate sensible choice, setting the $Q$ axis to the polarization angle of the
$A$ or $B$ detector, produces similar results in practice. 

The two detectors in an ideal orthogonally polarized 
pair would have the same $B_T$ as each other, 
the same $B_Q$ but with opposite sign from each other, and zero $B_U$. 
For that case, the sum of the two detectors in the pair (the ``pair sum'') has 
response to $T$ only and the difference between the two detectors in the pair 
(the ``pair difference'') has response only to $Q$. Imperfectly
matched $B_T$ between detectors in a pair 
causes temperature to polarization leakage, which is covered 
extensively in Section \ref{sec:diffbeam} and the Systematics Paper. In this 
Section, we focus on $B_Q$ and $B_U$.

The $B_T$, $B_Q$, and $B_U$ for each detector are measured from RPS observations. We start with
a set of beam maps for each detector, taken at 24~RPS grid angles spaced by
$15\deg$~steps. First, the maps are rescaled to correct for the collimation
offset described in Section~\ref{sec:rps} and to achieve uniform relative 
calibration for paired detectors. Then, the 24 maps for each detector are
summed with uniform weighting to form $T$~maps, $\cos 2\theta$ weighting to
form $Q$ maps, and $\sin 2\theta$ weighting to form $U$ maps, where $\theta$ is
equal to the angle between the RPS polarization axis and the detector $Q$ axis.

\begin{figure}[ht]
   \begin{center}
   \begin{tabular}{c}
   \includegraphics[height=10cm]{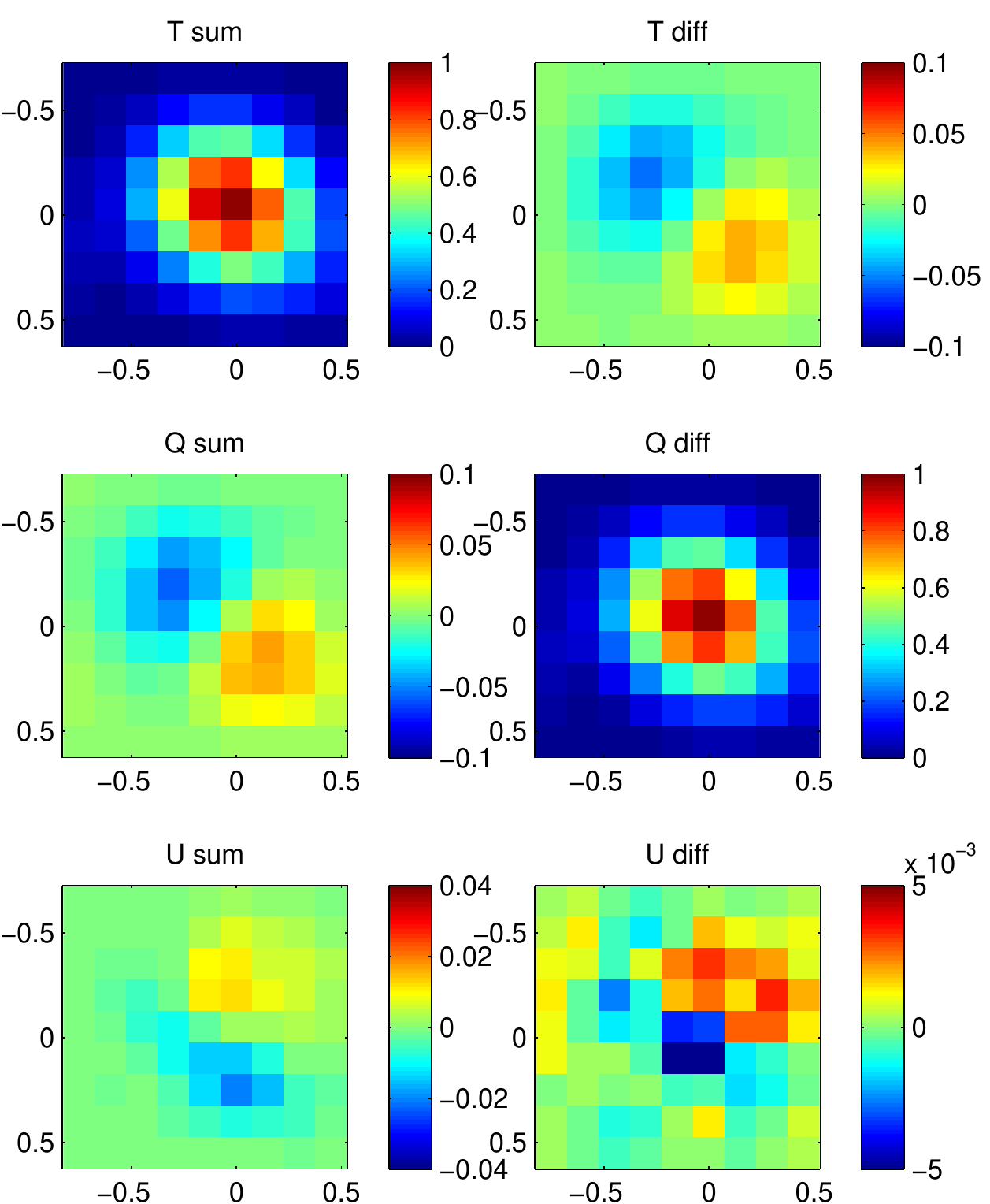}
   \end{tabular}
   \end{center}
   \caption{ \label{fig:crosspol}  
The Stokes $T$, $Q$, and $U$ beam maps ($B_T$, $B_Q$, and $B_U$)
for a single typical pixel in \bicepp from RPS measurements, smoothed
with a $0.1\degr$ Gaussian kernel.
The left column shows the response of the sum of the detectors in a pair; the right column shows the 
pair difference response. The
pair difference $B_T$ and pair sum $B_Q$ both 
show the differential pointing present in \bicepp.
An ideal instrument would have no $U$ response.
Only the pair difference beams are relevant to \bicepp polarization analysis.
The small ($\lesssim0.8\%$) features in the pair difference $B_U$
cause a negligible amount of E-to-B leakage.
The larger feature in the pair sum $B_U$ beam would cause polarization to temperature 
leakage, which is harmless.  
Note that the color scales are not uniform across panels.
   } 
   \end{figure}

Figure~\ref{fig:crosspol} shows the measured $B_T$, $B_Q$, and $B_U$ for the pair sum
and pair difference combinations of a single typical pixel in \bicepp. To 
reduce measurement noise, these maps have been smoothed with a $0.1\degr$
Gaussian kernel.
The pair difference $B_T$ (upper right panel) shows the differential pointing that is prominent in \bicepp.
Structure in the pair sum $B_Q$ and $B_U$ causes polarization to temperature 
leakage, which is unimportant.

While the monopole response of the pair difference $B_U$ is zero by 
construction, there is some higher-order response to $U$, 
which can be seen in the bottom panels of Figure~\ref{fig:crosspol}.
We do not include these higher-order $B_U$ effects in the main power spectrum analysis;
therefore, power in the pair difference $B_U$ could cause uncorrected polarization rotation that would lead 
to E-to-B leakage.
The amplitude of the pair difference $B_U$ is typically $\lesssim0.8\%$ of $B_Q$.
In the power spectrum this amplitude is squared, so the effective EE-to-BB leakage 
is $\lesssim 6\times10^{-5}$.
The resulting contamination at \mbox{$\ell\sim90$--125} is $\sim5\times10^{-5}\mu\textrm{K}^2$.
This is a factor of~10 below the $r=0.01$ BB signal, so this effect is negligible for \bicepp.
Furthermore, boresight rotation and variation among detectors can provide additional cancellation of 
the E-to-B leakage, 
although we do not rely on such cancellation in the above argument.

\section{Simulation and deprojection of mismatched elliptical beams}
\label{sec:simulation}

The \bicepp and \keck simulation pipeline is fully described
in the Systematics Paper and the Results Paper.
We give here a brief description, discussing the role of detailed beam map measurements
in simulations.

\begin{figure}[ht]
   \begin{center}
   \begin{tabular}{c}
   \includegraphics[height=6.5cm]{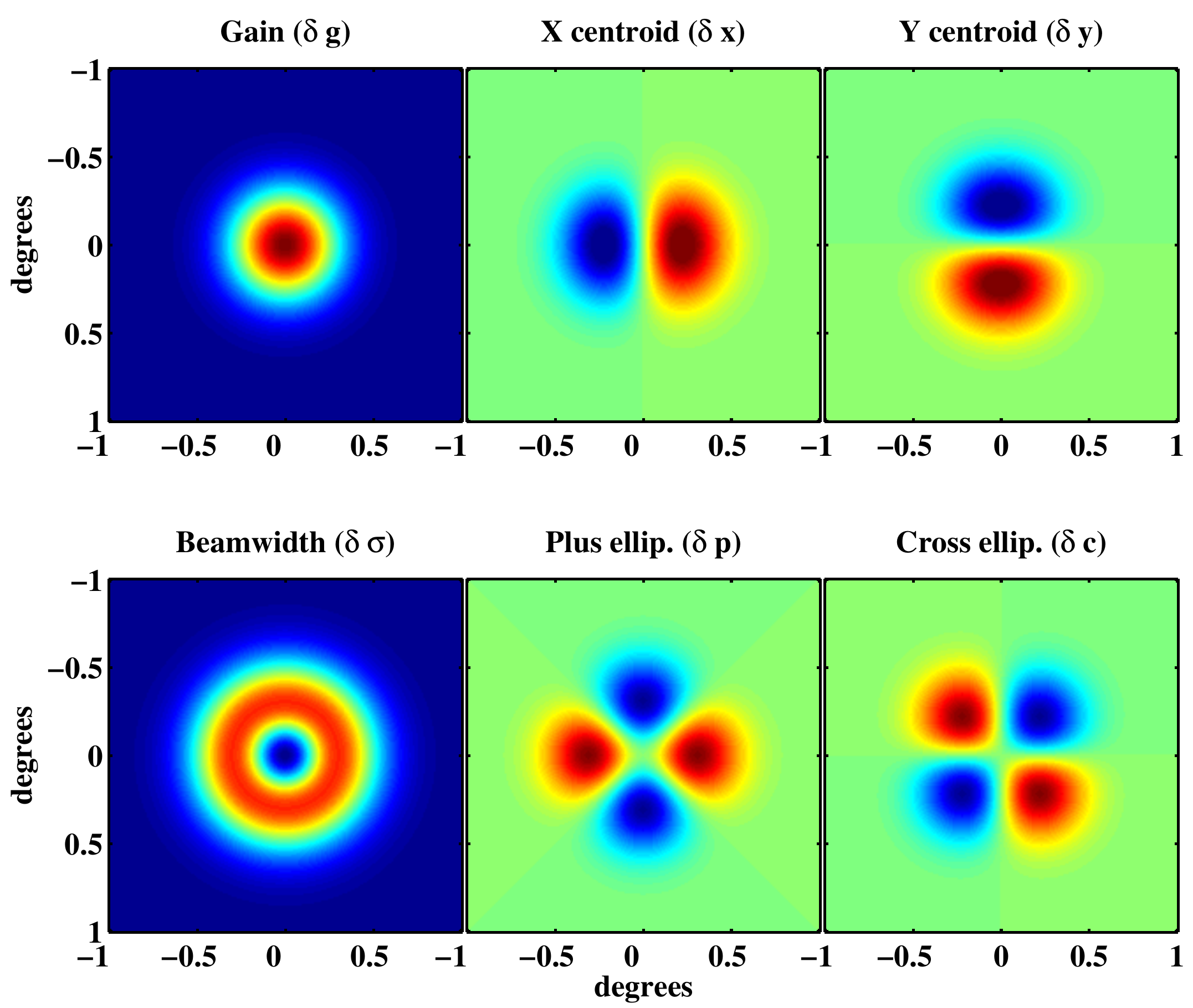}
   \end{tabular}
   \end{center}
   \caption{ \label{fig:diffBeam}  Differential beam templates resulting in mismatch in (a) responsivity 
(b) x-position (c) y-position (d) beam width (e) 
ellipticity in plus orientation (f) ellipticity in cross orientation.  In the limit of small differential 
parameters, a differenced beam 
pattern constructed from the difference of two elliptical Gaussians
can be represented as a linear combination of each of these templates. 
   } 
   \end{figure}

The beam pattern of a single detector can be characterized as a set of 
perturbations
on a circular Gaussian fit with a nominal width ($\sigma_{\mathrm{n}}$) 
equal to the receiver-averaged value and a nominal beam 
center equal to the calculated center for the pair of detectors from the initial elliptical Gaussian fit.
We consider the first six perturbations, corresponding
to the templates shown in Figure~\ref{fig:diffBeam}.  These six templates correspond to relative responsivity, 
x-position offset,
y-position offset, beam width, ellipticity in the ``plus'' orientation, and ellipticity in the ``cross'' orientation.
To first and second order, these six perturbations
of a circular Gaussian directly relate to the derivatives of the beam-convolved
temperature sky, $T$, which we use to remove temperature to polarization leakage.
The data processing pipeline is capable of removing leakage induced by beam mismatch modes that correspond to
mismatch modes of elliptical Gaussian beams, and is described in the Systematics Paper.
However, as shown in Figure~\ref{fig:farFieldBeam}, the beams are not perfectly
described by an elliptical Gaussian.
We take advantage of the detailed, high signal-to-noise
beam maps described in Section~\ref{sec:beamchar}
to fully describe the response of the detectors in the far field.
To capture the effects of each detector's beam on the 
science data and to understand any residual beam mismatch leakage after deprojection,
we use the beam maps to run ``beam map'' simulations.

The beam map simulations can use an arbitrary two-dimensional 
convolution kernel for the beam of each detector, which is convolved
with a flat projection of the 143~GHz Planck temperature map~\citep{planckI,planckVI} and interpolated
to produce simulated detector timestreams. The simulated timestreams are then fed
into the regular data processing pipeline, ensuring that
the processing and filtering of the real data are applied to
the simulated timestreams. 
The simulated timestreams are binned into
maps and then deprojected with the same template
that is used for real data.
The flat-sky approximation in principle limits the accuracy of
the simulation at a level of $r \sim 10^{-4}$, which in practice 
is lower than the observed leakage, discussed in the Systematics Paper. 

Since we have high-quality, high signal-to-noise 
beam maps for nearly every detector used in the science analysis,
we use these beam maps as two-dimensional convolution kernels in the beam map
simulations, allowing us to study precisely the effects of the 
real beam on the science data for every detector.  The results of the beam map
simulations have been used in the Results Paper.

   \begin{figure*}[ht]
   \begin{center}
   \begin{tabular}{c}
     \includegraphics[height=12cm]{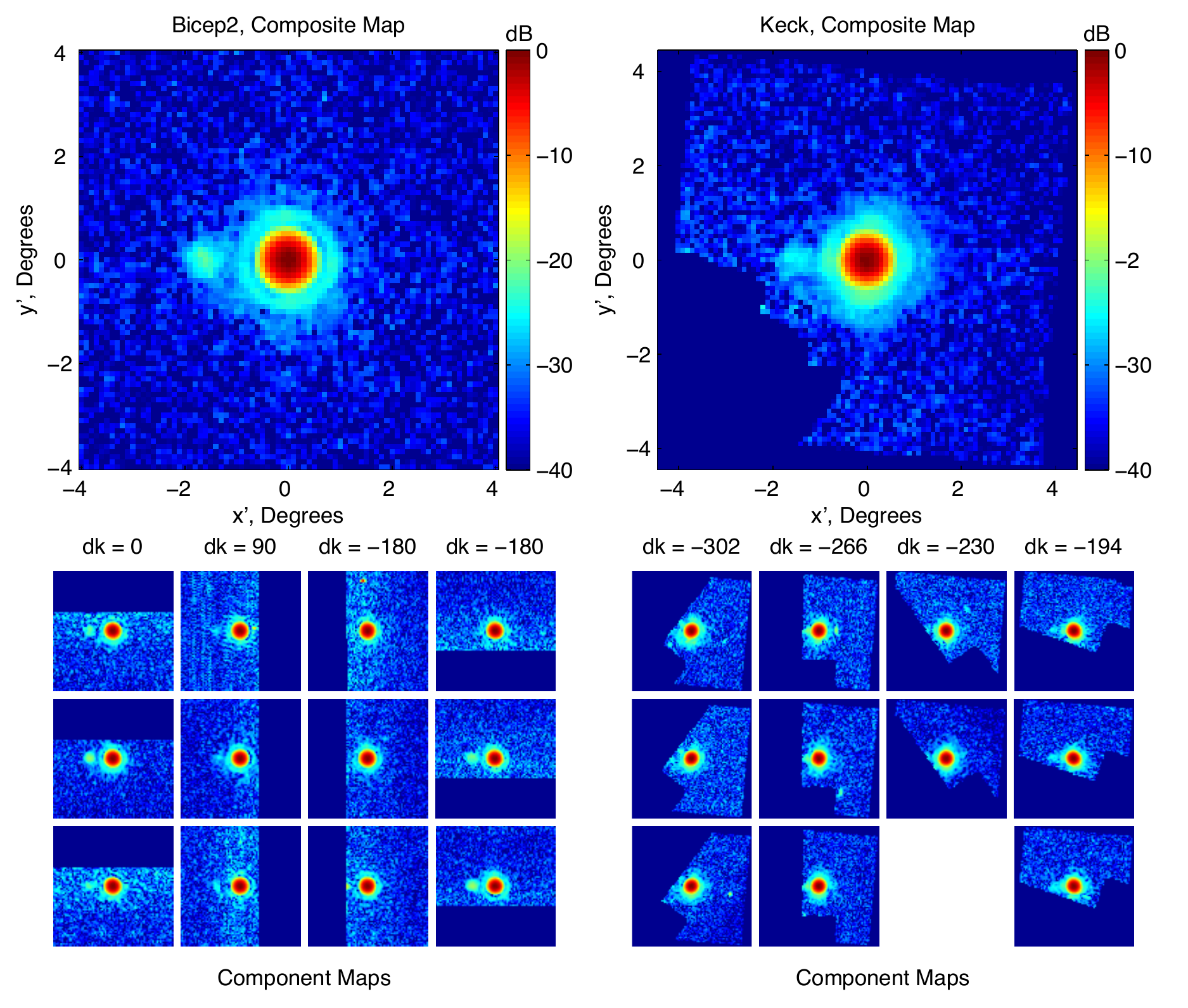}
   \end{tabular}
   \end{center}
   \caption{ \label{fig:compositebm} 
     Left: An example composite beam map for a \bicepp detector using twelve beam maps. 
     Right: An example composite beam map
     for a \keck detector using eleven beam maps.  The maps are rotated to account
     for the boresight rotation angle and then added.  The color scale is logarithmic with decades marked in dB.
   } 
   \end{figure*}

\subsection{Per-detector beam maps}
The two-dimensional convolution kernels that are fed into the beam map 
simulation pipeline are constructed from a composite of the the far-field beam maps 
that we describe in Section~\ref{sec:beamchar}.
For \bicepp, we use beam maps taken with the 45~cm 
diameter thermal source. The maps were taken in November 2012,
and the complete set consists of a total of twelve maps:
three sets of maps taken at four boresight rotation angles
separated by $90^\circ$. The three different sets of maps were taken
using different aluminum transition bias points to 
find the optimal response for as many detectors as possible and to
reduce gain compression artifacts that appeared in a small subset of detectors.
The thermal source is only $\sim2^\circ$ above the horizon, so
we mask out the ground in these beam maps by masking the portion of the map
that is $>1.5^\circ$ from the main beam and along the horizon. Even with a chopped
source, the hot ground causes response in the detectors that is visible in the 
demodulated maps.
We rotate the maps to account for the boresight angle, 
and the maps are then centered on the common beam centroid
for each detector pair.  To make the composite map that is used in simulations, 
we take the median amplitude for each pixel across all the component maps.
The left-hand panel of Figure~\ref{fig:compositebm} shows an example of the composite beam map
for a single detector that has been built from the set of 
twelve component maps for \bicepp.

Taking a median filter across all maps allows us to downweight
spurious signals in the individual maps that are not repeatable
across maps, allowing us to make clean, high signal-to-noise maps.
We find that the level of the noise in the composite beam maps
is low enough to show no effect in the beam map simulations.
Due to masking and rotation, 
not all pixels in the composite beam map use the same number of input beam maps.
For a radius of $<1.2^\circ$ from the beam center, all twelve component maps
are included in the composite map, and all pixels use at least three 
beam maps.

The \keck composite beam maps for 2012 and 2013 are constructed from
sets of maps taken in February 2012 and February 2013 respectively.
The maps taken in February 2012 use the 20~cm diameter thermal source,
and as a result have lower signal levels than the \bicepp component
maps and the \keck February 2013 maps, which use the 45~cm diameter
thermal source. 
The \keck beam maps are constructed from a set of $\sim25$ beam maps taken
at ten different boresight rotation angles. 
For each receiver, we use beam maps taken at up to five boresight rotation angles
where the receiver was positioned so that its beams reflect off of the large flat
mirror and to the thermal source.  The drum must be rotated so that a given receiver is
physically near the bottom of the drum for the beams from that receiver to reflect off
of the flat mirror and to the source.  Therefore, only boresight rotation angles that place a
given receiver near the bottom of the drum are used for composite maps for that receiver.

For each of the five boresight rotation angles for which detectors in a given receiver
view the mirror, we cut maps for pixels that do not see the mirror. 
The number of maps that are included in each detector's composite map
for the \keck varies from a single map (for detectors at the edge of a focal plane
on Receiver~0 for 2013)
to a maximum of eleven component maps with a median of nine maps used per detector.
In addition to masking the ground in \keck beam maps, we also
mask the South Pole Telescope by masking out a rectangle that is
$5\deg$ wide and $2.5\deg$ high, beginning $2.5\deg$ to the left of the chopped thermal 
source as viewed by \keck receivers.  We do not have beam maps for 9~detector pairs 
used in \keck analysis for~2012 and 28~detector pairs used in \keck analysis for~2013.

The right-hand panel of Figure~\ref{fig:compositebm}
shows an example of the resulting composite beam map for a single
detector for the \keck. As a result of the limited 
boresight rotation angle coverage, there is a portion of each map that is masked out in the composite
map for each detector. 
We fill in the unmapped parts of the composite beam map by
inserting the mean value from an azimuthal average around the beam center.

\subsection{Simulation Results}

   \begin{figure*}[ht]
   \begin{center}
   \begin{tabular}{c}
   \includegraphics[width=17.5cm]{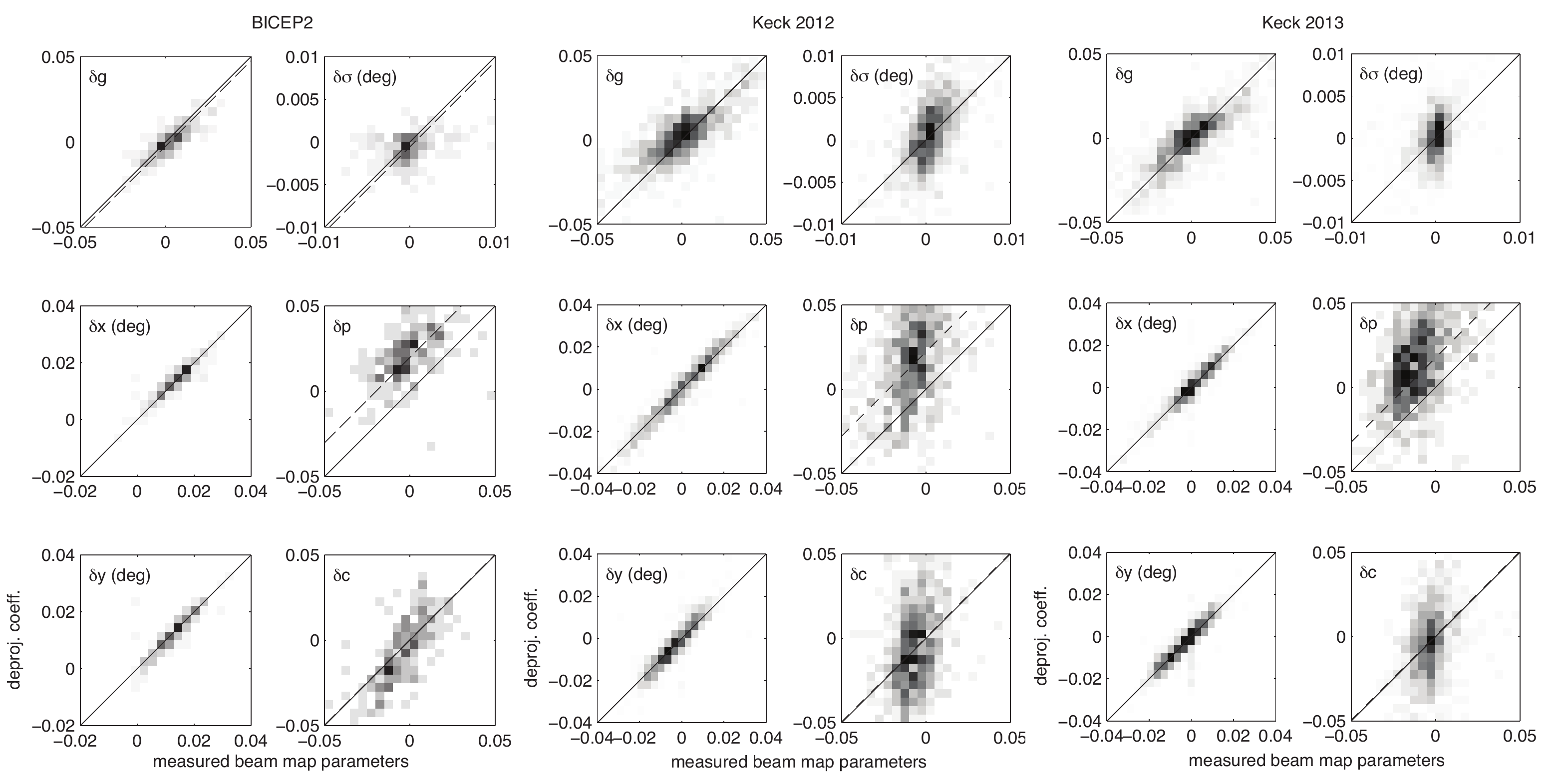}
   \end{tabular}
   \end{center}
   \caption{ \label{fig:deprojVsBeamMap}  A comparison of the deprojection coefficients recovered using our 
     CMB observation data to the measured beam parameters for \bicepp and 
     the \keck 2012 and 2013.  Measured differential gains for \bicepp and the \keck  
     are determined using the cross-correlation of 
     T maps for individual detectors with Planck.  The rest of the measured beam parameters are from
     beam maps.
     We observe a strong correlation for differential pointing and differential ellipticity for \bicepp, and a 
     strong correlation for differential pointing in the \keck. The
     scatter for differential ellipticity 
     for the \keck is higher than for \bicepp because we have
     less data for the \keck compared to \bicepp so the noise level is higher for the coefficients recovered from
     CMB observation data.  The solid line indicates a one-to-one correlation.  
     The bias in the recovered deprojection coefficients 
     predicted by simulations is shown with the dashed line as an offset and is discussed in the Systematics Paper.} 
   \end{figure*}

Comparing the beam map simulations
with regular simulations, where beam mismatch modes are derived from CMB
temperature data itself, we find that
the beam map simulations accurately predict the leakage
of the beam mismatch modes, discussed
fully in the Systematics Paper.
When we do not deproject any main beam mismatch
modes, the contamination of the B-mode 
auto-spectrum is very well predicted by the 
beam map simulation.
The beam map simulation spectra also predict the jackknife failures
that we see in real data before deprojection.

Figure~\ref{fig:deprojVsBeamMap}
compares the deprojection coefficients derived using
CMB data to the measured beam parameters
for \bicepp and the \keck for the 2012 and 2013 configurations.
Measured differential gains for \bicepp and the \keck 
are determined using the cross-correlation of 
T maps for individual detectors with Planck.  The rest of the measured beam parameters are from
beam maps.  The deprojection coefficients for \bicepp
show a correlation for all main beam mismatch modes,
consistent with the observation that
the measured beam parameters are
the same main beam mismatch modes that are present in the real data.
We also see a strong correlation for the differential pointing mismatch modes for the \keck.
The \keck differential gain, beam width, and ellipticity modes
show a large scatter in the recovered deprojection
coefficients from real data compared to the 
measured beam parameters.  This 
is because the deprojection coefficients for the \keck
are obtained from one year of CMB data (since 2012 and 2013 must be calculated separately due to different 
receiver configurations), 
compared to three years for \bicepp, resulting
in higher noise levels in the deprojection coefficients derived from real data.

\subsection{Undeprojected residual mismatch}
\label{sec:residual}

As described in the Systematics Paper, for the \bicepp B-mode analysis presented in the Results Paper,
we deproject differential pointing and gain and subtract the effects of differential
ellipticity.  These modes, however, do not fully describe the
beams of \bicepp and the \keck, which have small contributions from higher-order terms.  
Figure~\ref{fig:farFieldBeam} shows an example detector and the residual power after
removing an elliptical Gaussian fit, displaying the power contained in
higher-order terms.  The power in the per-pair difference beam 
that is not described by the six parameters in Figure~\ref{fig:diffBeam} could 
be a source of temperature to polarization leakage.

Using beam map simulations, we 
can predict the amount of contamination in our BB spectrum from these additional (higher-order) 
beam mismatch modes that we do not deproject.  Since the beam map simulations use
the measured far-field response of each beam as inputs,
they include the 
effect of all beam mismatch modes within $4^\circ$ of the beam center, limited only by the level of noise
in the measurement of the far-field beams.
Any effect from far sidelobes and residual power outside of $4^\circ$ from the beam center
have been shown to be small~(see Section~\ref{sec:farsidelobes}).

After deprojecting differential pointing, gain, and ellipticity, the level of contamination predicted
by beam map simulations is well below the sensitivity of \bicepp, as described in the 
Systematics Paper. 
For \bicepp, we 
have mitigated temperature to polarization leakage caused by per-pair beam mismatch
to a level sufficient to detect $r\simeq0.003$~\citep{systematics}.  We 
have also limited the contribution from additional systematics to $r\lesssim0.006$~\citep{systematics}.

\section{Conclusions}
We have fully described the optical system and 
characterized the optical performance of the \bicepp experiment 
and the \keck 2012 and 2013 configurations.  We have performed a full beam mapping campaign 
{\it in situ} at the South Pole and have measured far-field beam shape parameters, near-field
beam shapes, detector polarization angles, per-pair cross-polar response, and far-sidelobe response.

We find that measured beam shapes match physical optics simulations well, but that for a given pair
of orthogonally polarized detectors, there can be significant differences in beam shape in the far field, 
especially in differential pointing between the two detectors.  We find that the level of E-mode
to B-mode leakage is less than $r\simeq0.001$ for \bicepp, and in the Systematics Paper we
show that the remaining temperature to polarization leakage due to residual, 
higher-order components of the differential beam
is at the level of $r\simeq0.003$ for \bicepp, well below the sensitivity of the experiment.
We expect the level of leakage due to beam effects to be similar for the \keck.

\acknowledgements

\bicepp was supported by the National Science Foundation (NSF) under
grants ANT-0742818 and ANT-1044978 (Caltech/Harvard) and ANT-0742592
and ANT-1110087 (Chicago/Minnesota).  The \keck was supported by the NSF
under grants ANT-1145172 (Harvard), ANT-1145143 (Minnesota), and ANT-1145248 (Stanford),
and by the W. M. Keck Foundation (Caltech).
The development of antenna-coupled
detector technology was supported by the JPL Research and Technology
Development Fund and grants 06-ARPA206-0040 and 10-SAT10-0017
from the NASA APRA and SAT programs.  The development and testing of
focal planes were supported by the Gordon and Betty Moore Foundation
at Caltech.  Readout electronics were supported by a Canada Foundation
for Innovation grant to UBC.  
Computations presented in this paper were run on the Odyssey cluster supported 
by the FAS Science Division Research Computing Group at Harvard.  The analysis 
effort at Stanford and SLAC was partially suported by the U.S. Department of Energy
Office of Science.
The receiver development was supported
in part by a grant from the W. M. Keck Foundation.
Tireless administrative support was provided by
Irene Coyle and Kathy Deniston.

We thank the staff of the US Antarctic Program and in particular
the South Pole Station without whose help this research would not have been possible.
Most special thanks go to our heroic winter-overs Robert Schwarz and Steffen Richter.
We thank all those who have contributed past efforts to the \bicep /\keck\
series of experiments, including the \bicepo team,
as well as our colleagues on the \spider team with whom
we coordinated receiver and detector development efforts at Caltech.

\bibliographystyle{apj}
\bibliography{beams}

\begin{thebibliography}{}
\expandafter\ifx\csname natexlab\endcsname\relax\def\natexlab#1{#1}\fi

\bibitem[{{Ade} {et~al.}(2006){Ade}, {Pisano}, {Tucker}, \& {Weaver}}]{ade06}
{Ade}, P.~A.~R., {Pisano}, G., {Tucker}, C., \& {Weaver}, S. 2006, in Society
  of Photo-Optical Instrumentation Engineers (SPIE) Conference Series, Vol.
  6275, Society of Photo-Optical Instrumentation Engineers (SPIE) Conference
  Series

\bibitem[{{Ahmed} {et~al.}(2014){Ahmed}, {Amiri}, {Benton}, {Bock},
  {Bowens-Rubin}, {Buder}, {Bullock}, {Connors}, {Filippini}, {Grayson},
  {Halpern}, {Hilton}, {Hristov}, {Hui}, {Irwin}, {Kang}, {Karkare}, {Karpel},
  {Kovac}, {Kuo}, {Netterfield}, {Nguyen}, {O'Brient}, {Ogburn}, {Pryke},
  {Reintsema}, {Richter}, {Thompson}, {Turner}, {Vieregg}, {Wu}, \&
  {Yoon}}]{bicep3}
{Ahmed}, Z., {Amiri}, M., {Benton}, S.~J., {et~al.} 2014, in Society of
  Photo-Optical Instrumentation Engineers (SPIE) Conference Series, Vol. 9153,
  Society of Photo-Optical Instrumentation Engineers (SPIE) Conference Series

\bibitem[{{Aikin}(2013)}]{aikinThesis}
{Aikin}, R.~W. 2013, PhD thesis, California Institude of Technology

\bibitem[{{Aikin} {et~al.}(2010)}]{aikin10}
{Aikin}, R.~W., {et~al.} 2010, Millimeter, Submillimeter, and Far-Infrared
  Detectors and Instrumentation for Astronomy V, 7741, 77410V

\bibitem[{{BICEP2 Collaboration}(2014{\natexlab{a}})}]{instrument}
{BICEP2 Collaboration}. 2014{\natexlab{a}}, The Astrophysical Journal, 792, 62

\bibitem[{{BICEP2 Collaboration}(2014{\natexlab{b}})}]{results}
---. 2014{\natexlab{b}}, Physical Review Letters, 112, 241101

\bibitem[{{BICEP2 Collaboration}(2015)}]{systematics}
---. 2015, arXiv:1502.00608

\bibitem[{{BICEP2, \keck, and \spider Collaborations}(2015)}]{detectors}
{BICEP2, \keck, and \spider Collaborations}. 2015, arXiv:1502:00619

\bibitem[{{Bradford}(2012)}]{bradford}
{Bradford}, K.~J. 2012, Undergraduate thesis, Harvard University

\bibitem[{Carroll {et~al.}(1990)Carroll, Field, \& Jackiw}]{carroll}
Carroll, S.~M., Field, G.~B., \& Jackiw, R. 1990, Phys. Rev. D, 41, 1231

\bibitem[{{Filippini} {et~al.}(2010){Filippini}, {Ade}, {Amiri}, {Benton},
  {Bihary}, {Bock}, {Bond}, {Bonetti}, {Bryan}, {Burger}, {Chiang}, {Contaldi},
  {Crill}, {Dor{\'e}}, {Farhang}, {Fissel}, {Gandilo}, {Golwala},
  {Gudmundsson}, {Halpern}, {Hasselfield}, {Hilton}, {Holmes}, {Hristov},
  {Irwin}, {Jones}, {Kuo}, {MacTavish}, {Mason}, {Montroy}, {Morford},
  {Netterfield}, {O'Dea}, {Rahlin}, {Reintsema}, {Ruhl}, {Runyan}, {Schenker},
  {Shariff}, {Soler}, {Trangsrud}, {Tucker}, {Tucker}, \& {Turner}}]{spider}
{Filippini}, J.~P., {Ade}, P.~A.~R., {Amiri}, M., {et~al.} 2010, in Society of
  Photo-Optical Instrumentation Engineers (SPIE) Conference Series, Vol. 7741,
  Society of Photo-Optical Instrumentation Engineers (SPIE) Conference Series

\bibitem[{{Hinderks} {et~al.}(2009){Hinderks}, {Ade}, {Bock}, {Bowden},
  {Brown}, {Cahill}, {Carlstrom}, {Castro}, {Church}, {Culverhouse},
  {Friedman}, {Ganga}, {Gear}, {Gupta}, {Harris}, {Haynes}, {Keating}, {Kovac},
  {Kirby}, {Lange}, {Leitch}, {Mallie}, {Melhuish}, {Memari}, {Murphy},
  {Orlando}, {Schwarz}, {Sullivan}, {Piccirillo}, {Pryke}, {Rajguru},
  {Rusholme}, {Taylor}, {Thompson}, {Tucker}, {Turner}, {Wu}, \&
  {Zemcov}}]{quad}
{Hinderks}, J.~R., {Ade}, P., {Bock}, J., {et~al.} 2009, Astrophys. J., 692,
  1221

\bibitem[{{Kamionkowski} {et~al.}(1997){Kamionkowski}, {Kosowsky}, \&
  {Stebbins}}]{kamionkowski97}
{Kamionkowski}, M., {Kosowsky}, A., \& {Stebbins}, A. 1997, Physical Review
  Letters, 78, 2058

\bibitem[{{Keating} {et~al.}(2013){Keating}, {Shimon}, \& {Yadav}}]{keating}
{Keating}, B.~G., {Shimon}, M., \& {Yadav}, A.~P.~S. 2013, Astrophys. J. Lett.,
  762, L23

\bibitem[{{Keating} {et~al.}(2003)}]{Keating03}
{Keating}, B.~G., {et~al.} 2003, Proc.\ SPIE Int.\ Soc.\ Opt.\ Eng., 4843, 284

\bibitem[{{Lamb}(1996)}]{lamb}
{Lamb}, J.~W. 1996, International Journal of Infrared and Millimeter Waves, 17,
  1997

\bibitem[{{Leitch} {et~al.}(2002){Leitch}, {Pryke}, {Halverson}, {Kovac},
  {Davidson}, {LaRoque}, {Schartman}, {Yamasaki}, {Carlstrom}, {Holzapfel},
  {Dragovan}, {Cartwright}, {Mason}, {Padin}, {Pearson}, {Readhead}, \&
  {Shepherd}}]{dasi}
{Leitch}, E.~M., {Pryke}, C., {Halverson}, N.~W., {et~al.} 2002, Astrophys. J.,
  568, 28

\bibitem[{{O'Brient} {et~al.}(2012){O'Brient}, {Ade}, {Ahmed}, {Aikin},
  {Amiri}, {Benton}, {Bischoff}, {Bock}, {Bonetti}, {Brevik}, {Burger},
  {Davis}, {Day}, {Dowell}, {Duband}, {Filippini}, {Fliescher}, {Golwala},
  {Grayson}, {Halpern}, {Hasselfield}, {Hilton}, {Hristov}, {Hui}, {Irwin},
  {Kernasovskiy}, {Kovac}, {Kuo}, {Leitch}, {Lueker}, {Megerian}, {Moncelsi},
  {Netterfield}, {Nguyen}, {Ogburn}, {Pryke}, {Reintsema}, {Ruhl}, {Runyan},
  {Schwarz}, {Sheehy}, {Staniszewski}, {Sudiwala}, {Teply}, {Tolan}, {Turner},
  {Tucker}, {Vieregg}, {Wiebe}, {Wilson}, {Wong}, {Wu}, \& {Yoon}}]{roger}
{O'Brient}, R., {Ade}, P.~A.~R., {Ahmed}, Z., {et~al.} 2012, in Society of
  Photo-Optical Instrumentation Engineers (SPIE) Conference Series, Vol. 8452,
  Society of Photo-Optical Instrumentation Engineers (SPIE) Conference Series

\bibitem[{{Ogburn IV} {et~al.}(2010)}]{ogburn10}
{Ogburn IV}, R.~W., {et~al.} 2010, Proceedings of the SPIE: Millimeter,
  Submillimeter, and Far-Infrared Detectors and Instrumentation for Astronomy
  V, 7741, 77411G

\bibitem[{{Planck Collaboration}(2014{\natexlab{a}})}]{planckI}
{Planck Collaboration}. 2014{\natexlab{a}}, Astr. \& Astroph., 571, A1

\bibitem[{{Planck Collaboration}(2014{\natexlab{b}})}]{planckXXII}
---. 2014{\natexlab{b}}, Astr. \& Astroph., 571, A22

\bibitem[{{Planck HFI Core Team}(2011)}]{planckVI}
{Planck HFI Core Team}. 2011, Astr. \& Astroph., 536, A6

\bibitem[{{Runyan} {et~al.}(2003){Runyan}, {Ade}, {Bhatia}, {Bock}, {Daub},
  {Goldstein}, {Haynes}, {Holzapfel}, {Kuo}, {Lange}, {Leong}, {Lueker},
  {Newcomb}, {Peterson}, {Reichardt}, {Ruhl}, {Sirbi}, {Torbet}, {Tucker},
  {Turner}, \& {Woolsey}}]{runyan2003}
{Runyan}, M.~C., {Ade}, P.~A.~R., {Bhatia}, R.~S., {et~al.} 2003, Astrophys. J.
  Suppl., 149, 265

\bibitem[{{Seljak}(1997)}]{seljak97b}
{Seljak}, U. 1997, Astrophys. J., 482, 6

\bibitem[{{Seljak} \& {Zaldarriaga}(1997)}]{seljak97a}
{Seljak}, U., \& {Zaldarriaga}, M. 1997, Physical Review Letters, 78, 2054

\bibitem[{{Sheehy} {et~al.}(2010)}]{sheehy10}
{Sheehy}, C.~D., {et~al.} 2010, Millimeter, Submillimeter, and Far-Infrared
  Detectors and Instrumentation for Astronomy V, 7741, 77411R

\bibitem[{Takahashi {et~al.}(2010)}]{takahashi10}
Takahashi, Y.~D., {et~al.} 2010, Astrophys. J., 711, 1141

\bibitem[{Vieregg {et~al.}(2012)}]{vieregg12}
Vieregg, A.~G., {et~al.} 2012, Proceedings of the SPIE: Millimeter and
  Submillimeter Detectors and Instrumentation for Astronomy VI, 8452

\bibitem[{{Yoon} {et~al.}(2006)}]{yoon06}
{Yoon}, K.~W., {et~al.} 2006, in Society of Photo-Optical Instrumentation
  Engineers (SPIE) Conference Series, Vol. 6275, Society of Photo-Optical
  Instrumentation Engineers (SPIE) Conference Series

\end{thebibliography}

\end{document}